&&
\documentstyle[aps,eqsecnum,preprint,epsf]{revtex}
\begin{document}
\renewcommand{\thesection}{\arabic{section}}
\draft
\title{ \hfill   {\rm Submitted 23rd March 1995 }\\~~\\
                 Anomalous Scaling in a Model
                 of Passive Scalar Advection:
Exact Results  }
\author {Adrienne L. Fairhall\cite{procaccia},Omri Gat\cite{procaccia},
Victor L'vov\cite{lvov}, Itamar Procaccia\cite{procaccia} }
\address{Departments of~~\cite{procaccia}Chemical Physics {\rm and}
{}~~\cite{lvov}Physics of Complex Systems,\\
 The Weizmann Institute of Science,
Rehovot 76100, Israel,\\
\cite{lvov}Institute of Automation and Electrometry,
 Ac. Sci. of Russia, 630090, Novosibirsk, Russia
  }
\maketitle
\widetext
\begin{abstract}
\leftskip 54.8pt
\rightskip 54.8pt
Kraichnan's model of passive scalar advection in which the
driving velocity field has fast temporal decorrelation is
studied as a case model  for understanding the appearance of
anomalous scaling in  turbulent systems. We demonstrate how the
techniques of renormalized perturbation theory lead (after exact
resummations) to equations for the  statistical quantities that
reveal also non perturbative effects. It is  shown that
ultraviolet divergences in the diagrammatic expansion translate
into anomalous scaling with the inner length acting as the
renormalization  scale. In this paper we compute analytically
the infinite set of anomalous exponents that stem from the
ultraviolet divergences. Notwithstanding, non-perturbative
effects furnish a possibility of  anomalous scaling based on the
outer renormalization scale. The mechanism  for this intricate
behavior is examined and explained in detail.  We show  that in
the language of L'vov, Procaccia and Fairhall [Phys. Rev. E {\bf
50}, 4684 (1994)] the problem is ``critical" i.e. the anomalous
exponent of  the scalar primary field $\Delta=\Delta_c$. This is
precisely the condition that allows for anomalous scaling in the
structure functions as well, and we prove that this anomaly
must be based on the outer renormalization  scale. Finally, we
derive the scaling laws that were proposed by Kraichnan for
this problem, and show that his scaling exponents are consistent
with our  theory.

\end{abstract} \leftskip 54.8pt
  \narrowtext

\section{Introduction} \label{intro}
The model of passive scalar advection  with rapidly
decorrelating velocity field which was introduced some time  ago
by R.H. Kraichnan \cite{68Kra} was suggested recently
\cite{94Kra} as a case model for understanding multiscaling in
the statistical description of nonlinear field theories. The
model is for a scalar field $\Theta({\bf r},t)$ where ${\bf r}$
is a point in $R^d$. This field satisfies the equation of motion

\begin{equation}
\label{advect}
	\big[\partial_t  - \kappa \nabla^2  +
	{\bf u}({\bf r},t) \cdot \bbox{\nabla}\big]
	 \Theta({\bf r},t) = f({\bf r},t).
\end{equation}
In this equation $f({\bf r},t)$ is the forcing and ${\bf
u}({\bf r},t)$ is the velocity field which is taken to be a
stochastic field, rapidly varying in time. The forcing is
taken to be delta correlated in time and space-homogeneous,
\begin{equation}
	\langle  f({\bf r},t)  f({\bf r}',t') \rangle =
	\Phi_0({\bf r}-{\bf r}')  \delta(t - t').
\end{equation}
We will study this model in the limit of large Peclet (Pe)
number, which is defined as the dimensionless ratio $U_L
L/\kappa$, where $U_L$ is the scale of the velocity fluctuations
on the outer scale $L$ of the system.  The most important
property of the driving velocity field from the point of
view of the scaling properties of the passive scalar is the
tensor ``eddy diffusivity" \cite{68Kra}
\begin{eqnarray}
\label{eddy_diff}
	h_{ij}({\bf R}) & \equiv & \int^\infty_0 d\tau
	\langle [u_i({\bf r}+{\bf R},t + \tau) - u_i({\bf r},t+ \tau)]
	\nonumber \\
	& \times & [u_j({\bf r}+{\bf R},t) - u_j({\bf r},t)] \rangle)
\end{eqnarray}
where the symbol $\langle ... \rangle$ in
Eq.(\ref{eddy_diff}) stands for an ensemble average with
respect to the statistics of ${\bf u}$ which is given {\em
a priori}. The scaling properties of the scalar depend
sensitively on the scaling exponent $\zeta_h$ that
characterizes the $R$ dependence of $h_{ij}({\bf R})$:
\begin{eqnarray}
        h_{ij}({\bf R}) &=  & h(R)
\left[(\zeta_h +1 ) \delta_{ij} -
        \zeta_h \frac{R_i R_j}{R^2}\right] \,,
\label{h_scal}  \\
h(R)& = & H \left (\frac{R}{\cal L} \right)^{\zeta_h} \ ,
\label{def-h}
\end{eqnarray}
where $\cal L$ is some characteristic scale of the driving
velocity field. The structure of the tensor $h_{ij}$  is
dictated by the incompressibility of the velocity field, and $H$
is a free dimensional parameter. By ``scaling properties" we
mean how various correlation and response functions of
$\Theta({\bf r},t)$ and its gradients depend on separation
distances. For example the structure functions of $\Theta({\bf
r},t)$ are
\begin{equation}
\label{s2n_def}
	S_{2n}({\bf R})   \equiv
	\langle [ \Theta({\bf r}+{\bf R}, t) -
	\Theta({\bf r}, t)]^{2n} \rangle \ .
\end{equation}
In writing this equation we assumed that the statistics of the
velocity field leads to a stationary and space homogeneous
ensemble of the scalar $\Theta$. If the statistics is also
isotropic, then $S_{2n}$  becomes a function of $R$ only,
independent of the direction of ${\bf R}$.  The scaling exponents of
the structure functions $S_{2n}(R)$ characterize their $R$
dependence in the limit of large Pe,
\begin{equation}
\label{zeta_def}
 	S_{2n}(R) \sim R^{\zeta_{2n}} ,
\end{equation}
when $R$ is in the ``inertial" interval of scales that will
be discussed later in this paper.

In Ref.\cite{94Kra} Kraichnan showed that when the  driving
velocity field  ${\bf u}({\bf r},t)$ is delta correlated in time
one can derive the exact  ``balance" equations for the
structure functions $S_{2n}(R)$:
\begin{equation}
	-\hat B(R)S_{2n}(R) = J_{2n}(R)\ .
\label{balance}
\end{equation}
In this equation $\hat B(R)$ is the linear operator that will be
used below repeatedly:
\begin{equation}
	\hat B(R)= - R^{1-d}{\partial \over \partial R} R^{d-1}h(R)
	{\partial \over \partial R}   \,,
\label{B}
\end{equation}
with $d$ being the space dimension. On the RHS of the balance
equations we have
\begin{equation}
	J_{2n}(R)=4n \kappa \left<
	[\Theta({\bf r}+{\bf R})- \Theta({\bf r})]^{2n-2}
	|\nabla \Theta({\bf r})|^2 \right> \ .
\label{J}
\end{equation}
Kraichnan conjectured that the  scaling dependence of
$J_{2n}(R)$ on $R$ when $R$ is in the inertial range is given
by the law
\begin{equation}
	J_{2n}(R)=
	J_2(R) S_{2n}(R)/S_2(R) \sim R^{\zeta_{2n}-\zeta_2}  \ .
 \label{krconj}
 \end{equation}
This scaling law led Kraichnan to far reaching conclusions
regarding the  scaling exponents $\zeta_{2n}$. Once inserted in
the balance equations this  conjecture resulted in the
prediction that the scaling exponents satisfy  the equation
\begin{equation}
\label{kr_screl}
\zeta_{2n}(\zeta_{2n}-
        \zeta_2 +d) = n d \zeta_2 \ . 
 \end{equation}
If so, this model may be the first nonlinear non-equilibrium
case where  ``multiscaling" can be explicitly demonstrated.
Further, Kraichnan et al. proposed in \cite{95KYC} that if
this model {\em is} multiscaling, then (\ref{kr_screl}) is
the unique solution for $\zeta_{2n}$. We
will see that among  other things our considerations lead to
precisely the scaling law  $J_{2n}(R)\sim
R^{\zeta_{2n}-\zeta_2}$ as conjectured by Kraichnan.  We will
see how the possibility of multiscaling is indeed realised
due to the analytic structure of the theory.

In a previous work on this model \cite{94LPF} (referred to
hereafter as paper I) renormalized perturbation theory was
employed to study the scaling behavior.  It is appropriate to
present first a short summary of the results of this paper.

The main point of paper I was that various statistical
quantities exhibit anomalous exponents that stem from
ultraviolet divergences in their diagrammatic expansion. For
example there exists a quantity called the nonlinear Green's
function (that is defined in sec.2 and is a 4-point quantity)
and which is denoted as
${\cal G}_2(0|{\bf r}_1,{\bf r}_2,{\bf r}_3,{\bf r}_4)$. It was shown that for
$r_1\simeq r_2 \simeq r$, $r_3 \simeq r_4
\simeq R$, and  $R \gg r$  (cf. I,
Eqs.(4.11), (4.12) and (4.18))
\begin{equation}
      \frac{\partial}{\partial r_1}\frac{\partial}{\partial r_2}
        {\cal G}_2(0|{\bf r}_1,r_2,r_3,r_4) \sim r^{-\Delta}\ ,
\label{anom}
\end{equation}
where $\Delta$ is an anomalous exponent that characterizes the
leading  divergence when $r\to 0$.   Moreover, the value of this
exponent is  important in determining much of the scaling
behavior in this model. It  appears prominently in $J_{2n}$ and
also in the correlation function of  passive dissipation
fluctuations.  The dissipation field is defined here  as
\begin{equation}
	\epsilon(x) \equiv \kappa |\nabla \Theta(x)|^2,
\end{equation}
and the correlation function $K(R)$ is
\begin{equation}
	K(R) = \langle \epsilon({\bf r}+{\bf R},t)
	\epsilon({\bf r},t) \rangle -
	\langle\epsilon(x)\rangle^2\ . \label{K}
\end{equation}
It was shown in paper I that the ultraviolet divergences
resulted in a  dependence on the inner cutoff of the theory,
denoted as $\eta$ [and  defined below in Eq.  (\ref{def-eta})]
which is written as:
\begin{equation}
 	K(R)/\kappa^2 \sim \eta^{-2\Delta}.  \label{KR}
\end{equation}
It was also explained in paper I that the theory indicates that
if  $\Delta$ reaches the critical value $\Delta_{\rm c}=
\zeta_h$ special  considerations must be made. For $\Delta <
\zeta_h$ it was shown that the  perturbation theory for $S_4$
and higher order structure functions  converges order by order
both in the ultra-violet and in the infra-red  limits.
Accordingly, there is no external renormalization length scale
in  the theory and there is no (perturbative) mechanism to ruin
simple  scaling, i.e.  $\zeta_{2n}=n\zeta_n$. In this case the
correlation of  dissipation fluctuations can be shown to decay
in the inertial range of  scales as
\begin{equation}
 	K(R) \sim \langle\epsilon(x)\rangle^2
 	\left({\eta \over R}\right)^{2\zeta_h-2\Delta}  \ .
\label{aa}
\end{equation}
Indeed, as long as $\Delta < \zeta_h$ the correlation decays in
the  inertial range as it must for a mixing field. On the
contrary, if $\Delta  = \zeta_h$ Eq.(\ref{aa}) cannot continue
to hold since it predicts that  the correlations do not decay.
This is precisely where the need for  anomalous scaling of the
structure functions comes in. We will see below  that the value
of $\Delta$ is precisely $\zeta_h$, and we will get  instead of
Eq.(\ref{aa}) the following prediction:
\begin{equation}
 	K(R) \sim \langle\epsilon(x)\rangle^2\left({L\over
 	R}\right)^{2\zeta_2-\zeta_4}  \ .
\label{ab}
\end{equation}
Here $L$ is the outer renormalization scale for this model,
that in general is not identical with the outer scale $\cal L$
of Eq.(\ref{def-h}). This result is in agreement with
Kraichnan's conjectures. The detailed  explanation of how this
phenomenon takes place is one of the major aims of  this paper.
It is an important mechanism that indicates how at least in
this example ultraviolet divergences coupled with
nonperturbative effects  may conspire together to give at the
end anomalous corrections which are  carried by the outer
renormalization scale. Whether or not such a  mechanism operates
in other hydrodynamic systems will be discussed in  separate
publications. At any rate, the example treated here serves as a
powerful demonstration of the fact that renormalized
perturbation  theory as applied to field theories of the
hydrodynamic type allows one,  after proper resummations, to
capture subtle nonperturbative effects.

It needs to be stressed that the anomalous exponent $\Delta$
discussed  above is just the leading divergence associated with
scalar anomalous  fields. We will explain below that there
exists a full spectrum of  anomalous exponents which are
associated with the inner length, and they  have to do with
anomalous fields of different irreducible  representation of the
rotation group\cite{95LL-a}. We will compute below  analytically
the whole spectrum of these exponents. For the model at  hand
this is an easy task, but it serves to demonstrate the rich
scaling  properties of hydrodynamic systems. This rich scaling
structure has not been considered by the fluid mechanics
community until now.

The structure of this paper is as follows: section 2 is devoted
to the derivation of the differential equations satisfied by the
$n$-point time correlations functions and the $n$-point
simultaneous correlation functions, as well as the equations for
the 2-point and 4-point Green's functions. The derivation is
based on the diagrammatic expansion of paper I, and the main
objective of this section is to demonstrate that this technique
yields the exact equations, and that the resulting equations
contain aspects of the problem which are not perturbative. The
results of the calculations of this section are identical to
alternative derivations which are based on standard stochastic
methods. Readers who are not interested in the method of
derivation can start to read this paper from section 3 which
begins with a catalog of the equations that are analyzed in the
rest of the paper. Section 3 begins  with the exact solution of
the Green's function and the 2-point correlation function. These
solutions are important since they introduce the homogeneous
solutions of the operator $\hat{\cal B}(R)$ which then appears importantly
in all the solutions of the higher order quantities. Section 2 B
deals with the exact solution of the 2-point correlation
function. This solution depends on the nature of the forcing. We
are able to study in detail how the effects of nonisotropic
forcing on the large scales decay in the inertial range as the
scale of observation decreases. The exponents that govern the
"law of isotropization" are important in determining the scaling
behavior of correlation functions of anomalous fields whose
presentation under the symmetry groups is different from
scalars. Finally, in section 2 C we discuss the 4-point
nonlinear Green's function which allows us the evaluation of the
anomalous exponent $\Delta$. Section 4 is devoted to the
calculation of $K(R)$, $J_4(R)$, and other correlations that
expose non-scalar anomalous fields. The section is based on
analyzing the equation for ${\cal F}_4$, and the strategy is to
extract the leading divergence that is characterized by the
anomalous exponent $\Delta$. Section 5 presents a brief
calculation of $J_{2n}$ and section 6 collects all the result
together in order to compute the scaling exponents of
$S_{2n}(R)$. We show that there are only two possibilities, one
is simple scaling, and the other is anomalous scaling in
agreement with Kraichnan's conjectures. Simple scaling is
possible only if the dissipation field is not mixing, which
seems a nonphysical condition. Section 7 offers a summary of the
paper.

\section{Diagrammatic derivation of the equations for correlation and
response functions}

In this chapter we demonstrate that for this model the exact
resummation of the diagrammatic expansion for the various
$n$-point correlation and response functions of the theory
results in exact differential equations that contain both
perturbative and nonperturbative anomalies. The analysis is
based on paper I, in which the first step is the
Belinicher-L'vov transformation\cite{94LPF,87BL,95LP-b}. This is
done by allowing the center of the coordinate system to move
along the Lagrangian trajectory of a particular fluid point. The
reference point is at position ${\bf r}_0$ at time $t_0$. The
trajectory of this point with respect to ${\bf r}_0$ is
\begin{equation}
	\bbox{\rho}({\bf r}_0,t_0|t) =
	\int_{t_0}^t dt' \,
	{\bf u}({\bf r}_0 + \bbox{\rho}({\bf r}_0,t_0|t'),t').
\end{equation}
Let us denote the transformed variables as
\begin{eqnarray}
	T({\bf r}_0,t_0|{\bf r},t) & = &
	\Theta({\bf r}+\bbox{\rho}({\bf r}_0,t_0|t),t), \\
	{\bf v}({\bf r}_0,t_0|{\bf r},t) & = &
	{\bf u}({\bf r}+\bbox{\rho}({\bf r}_0,t_0|t),t), \\
	\phi({\bf r}_0,t_0|{\bf r},t) & = &
	f({\bf r}+ \bbox{\rho}({\bf r}_0,t_0|t),t)  .
\end{eqnarray}

In terms of these new variables the equation of motion
(\ref{advect}) reads

\begin{equation}
\label{advect_ql}
     \!	\left[ \partial_t \!-  \kappa \nabla^2 - \!
	{\bf w}({\bf r}_0,t_0|{\bf r},t)
       \!\cdot\! \bbox{\nabla} \right] T({\bf r}_0,t_0|{\bf r},t)
 	=  \phi({\bf r}_0,t_0|{\bf r},t),
\end{equation}
where
\begin{equation}
	{\bf w}({\bf r}_0,t_0|{\bf r},t) =
	{\bf v}({\bf r}_0,t_0|{\bf r},t) -
	{\bf v}({\bf r}_0,t_0|{\bf r}_0,t).
\end{equation}
We turn now to the discussion of the statistical quantities that are defined
in terms of these variables.
\subsection{The 2-point quantities}

The 2-point functions discussed here are the 2-point correlation
function
\begin{equation}
\label{f_def}
	{\cal F}({\bf r}_0|x_1,x_2) \equiv \langle T(x_0|x_1)
	T(x_0|x_2) \rangle
\end{equation}
and the Green's function that is defined as
\begin{equation}
	{\cal G}({\bf r} _0|x_1,x_2)  \equiv
	\left\langle \frac{\delta T(x_0|x_1)}
	{\delta \xi(x_0|x_2)} \right\rangle
	\Bigg|_{\xi \rightarrow 0}\ , \label{g_def}
\end{equation}
where the $d+1$ dimensional vector $x \equiv ({\bf r},t)$,
and $\xi$ is the Belinicher-L'vov transformation of a
forcing that is added to the RHS of Eq.(\ref{advect_ql}).
The dependence on $t_0$ disappeared from the LHS of
Eqs.(\ref{g_def}) because one can prove
\cite{95LL-a} that all the average quantities are time
translationally invariant.

In this subsection we present the derivation of the equations
for ${\cal  F}({\bf r}_0|x_1,x_2)$ and ${\cal G}({\bf
r}_0|x_1,x_2)$. In paper I we  derived the Dyson-Wyld equations
for these two point functions.  For $t >  0$ they read
\begin{eqnarray}
(\partial_t &-& \kappa \nabla^2)
{\cal G}({\bf r}_0|{\bf r}_1,{\bf r}_2,t) = \delta({\bf r}_1-{\bf r}_2)
\delta(t)     \nonumber \\
& + & \int d{\bf r}' \Sigma({\bf r}_0|{\bf
        r}_1,{\bf r}') G({\bf r}_0|{\bf r}',{\bf r}_2,t) \label{dyson}\\
     {\cal F}({\bf r}_0\!\!\! &|&\!\!\!{\bf r}_1,{\bf r}_2,t) = \int d{\bf
        r}'d{\bf r}''\int_0^\infty dt' {\cal G}({\bf r}_0|{\bf r}_1,{\bf
        r}',t+t') \nonumber \\ & \times & [\Phi_0({\bf r}',{\bf
        r}'')+\Phi({\bf r}',{\bf r}'')] {\cal G}({\bf r}_0|{\bf r}_2,{\bf
        r}'',t').  \label{wyld} \end{eqnarray}
For negative times the  Green's function is zero, and the
correlation function is symmetric to  inverting time $t
\rightarrow -t$ and the coordinates ${\bf r} \rightarrow  -{\bf
r}$ together. In the Dyson equation (\ref{dyson}) the mass
operator $\Sigma({\bf r}_0|{\bf r},{\bf r}')$ can be written
explicitly
\begin{equation}
\label{sigma}
	\Sigma({\bf r}_0|{\bf r},{\bf r}') =
	\frac{\partial }{\partial  r_i}
	{\cal G}({\bf r}_0|{\bf r},{\bf r}',t=0)
	H_{ij}({\bf r}_0|{\bf r},{\bf r}')
	\frac{\partial}{\partial r'_j}
\end{equation}
whereas in the Wyld equation (\ref{wyld}) the mass operator $\Phi$
takes the form
\begin{equation}
\label{phi}
	\Phi({\bf r}_0|{\bf r},{\bf r}') =
	H_{ij}({\bf r}_0|{\bf r},{\bf r}')
	\frac{\partial}{\partial r_i} \frac{\partial}{\partial r'_j}
	{\cal F}({\bf r}_0|{\bf r},{\bf r}',t=0).
\end{equation}
In (\ref{sigma}) and (\ref{phi})
\begin{equation}
\label{h_def}
	H_{ij}({\bf r}_0|{\bf r},{\bf r}') =
	\int^\infty_{-\infty} dt \langle
	w_i({\bf r}_0,t_0|{\bf r},t) w_j({\bf r}_0,t_0|{\bf r}',0)
	\rangle.
\end{equation}
In the case considered in this paper in which the velocity
field is delta correlated in time the expression (\ref{h_def})
can be related to the eddy diffusivity (\ref{eddy_diff}),
expressed in terms of Eulerian correlations as:
\begin{equation}
\label{h_euler}
	H_{ij}({\bf r}_0|{\bf r},{\bf r}') =
	h_{ij}({\bf r}-{\bf r}_0) +  h_{ij}({\bf r}'-{\bf r}_0)
	- h_{ij}({\bf r}-{\bf r}')  .
\end{equation}
That the mass operators (\ref{sigma}) and (\ref{phi}) have an
explicit form corresponding to the 1-loop diagram rather than
an infinite series representation stems from the fact that the
velocity field decorrelates on an infinitely short time scale.
This simple form of the mass operators will be lost if we
relax this fast decay of the velocity time correlation
functions.

Equations (\ref{dyson}) and (\ref{wyld}) can be turned into
differential equations for the 2-point functions ${\cal F}$ and
${\cal G}$. Substituting (\ref{sigma}) in (\ref{dyson}) we
find that

\begin{eqnarray}
&&(\partial_t - \kappa\nabla^2)
	{\cal G}({\bf r}_0|{\bf r},{\bf r}',t)  =
	\delta({\bf r}-{\bf r}') \delta(t)
\label{g_eqmot}\\
\!    & +&\! \int  \! d{\bf r}''  {\partial \over \partial r_i}
    {\cal G}({\bf r}_0|{\bf r},{\bf r}'',0)
    H_{ij}({\bf r}_0|{\bf r},{\bf r}'')
    {\partial \over \partial r''_j}
      G({\bf r}_0|{\bf r}'',{\bf r}',t).
\nonumber
\end{eqnarray}
Remember that ${\cal G}({\bf r}_0|{\bf r},{\bf r}',t)$ is zero
for negative times. Integrating (\ref{g_eqmot}) over time from
$t=-\delta$ to $t=\delta$ and taking the limit $\delta
\rightarrow 0$ one finds that
\begin{equation}
	{\cal G}({\bf r}_0|{\bf r},{\bf r}',t=0^+) =
	\delta({\bf r}-{\bf r}')  \ .
\end{equation}
We will choose symmetrical regularization at $t=0$ and by
convention write
\begin{equation}
\label{g0}
 	{\cal G}({\bf r}_0|{\bf r},{\bf r}',t=0) = \frac{1}{2}
 	\delta({\bf r}-{\bf r}') .
\end{equation}
Using this evaluation the integration may be performed leading
to
\begin{equation}
\label{g_eq}
	\left[\partial_t + \hat{\cal D}_1({\bf r}-{\bf r}_0)
 \right]
	{\cal G}({\bf r}_0|{\bf r},{\bf r}',t)  =
	\delta({\bf r}-{\bf r}')\delta(t).
\end{equation}
Here we introduced the generalized  diffusion operator
\begin{equation}
        \hat{\cal D}_1({\bf r}) = - \kappa\nabla^2 +
       \hat {\cal B}({\bf r})  \ ,
\label{def_d1}
\end{equation}
where the operator $\hat{\cal B}$ is a key operator that appears
repeatedly  below. We will distinguish  between an operator
$\hat{\cal B} ({\bf r} _\alpha,{\bf r} _\beta)$ which acts on functions of
two variables ${\bf r} _\alpha$ and  ${\bf r} _\beta$ and an operator
$\hat{\cal B}({\bf R} )$ acting on functions of one variable ${\bf R} $:
\begin{eqnarray}
	\hat{\cal B}({\bf r} _\alpha,{\bf r}_\beta) &\equiv&
	\hat{\cal B}_{\alpha,\beta} =
	 h_{ij}({\bf r}_\alpha -{\bf r}_\beta) {\partial^2 \over
	\partial r_{\alpha,i} \partial r_{\beta,j} }  \,,
\label{def-B2}   \\
	\hat{\cal B}({\bf R}) &=& -h_{ij}(R) \frac{\partial^2}{\partial
     R_i \partial R_j   }  \ .
\label{def-B1}
\end{eqnarray}
Clearly  $\hat{\cal B}({\bf r}_\alpha,{\bf r}_\beta)$ turns into  $\hat
{\cal  B}({\bf R})$ on the class of functions depending on the
difference  ${\bf R} = {\bf r}_\alpha -{\bf r}_\beta$ only. In  spherical
coordinates the $\hat{\cal B} $-operator can be represented as the sum of
two contributions:
\begin{equation}
	\hat{\cal B}({\bf R}) = \hat B(R) +
	{(\zeta_h +d-1)\over (d-1)}{h(R) \over R^2} \hat L^2 \ .
\label{sep-B}
\end{equation}
Here $\hat B$ is the operator (\ref{B}) and  $\hat L$ is the
angular momentum operator $-i {\bf r}\times \bbox{\nabla}$
which depends only on the direction of ${\bf r}$.

For future purposes we also need the equation of the
time-integrated Green's function
${\cal G} ({\bf r}-{\bf r}_0 ,{\bf r}' -{\bf r}_0 )$
which is defined as
\begin{equation}
	{\cal G} ({\bf r}-{\bf r}_0 ,{\bf r}' -{\bf r}_0 ) =
	\int dt {\cal G}({\bf r}_0|{\bf r},{\bf r}',t)  .
\label{Gt}
\end{equation}
This function satisfies the equation
\begin{equation}
	\hat{\cal D}_1({\bf R}) {\cal G} ({\bf R} ,{\bf R}') =
	\delta ({\bf R} -{\bf R}') \ ,
\label{EqGt}
\end{equation}
which follows from Eq. (\ref{g_eq}).

The equation of motion (\ref{EqGt}) allows us to
introduce the inner scale  of this model, denoted by
$\eta$. By definition $\eta$ is the scale for which
the advective term $\hat{\cal B}$ is of the order of
the dissipative term $\kappa \nabla^2$. Equating the
two terms we get
\begin{equation}
	\eta  \simeq \left( \frac{\kappa}{H} \right)^{1/\zeta_h}.
\label{def-eta}
\end{equation}

We proceed now to determine an equation of motion for
the two-point  correlator ${\cal F}({\bf r}_0|{\bf
r}_1,{\bf r}_2,t)$. It is clear from Eq. (\ref{g_eq})
that one may define an inverse operator for the
Green's function  ${\cal G}({\bf r}_0|{\bf r},{\bf
r}',t)$ according to \begin{equation}
\label{g_inv}
	{\cal G}_1^{-1}({\bf r}-{\bf r}_0,t) \equiv
	 \partial_t - \hat{\cal D}_1({\bf r}-{\bf r}_0)\ .
\end{equation}
Note from Eq. (\ref{g_eq}) that the equation of motion for
${\cal G}({\bf r}_0|{\bf r},{\bf r}',t)$ only depends on the
first coordinate ${\bf r}$, so that ${\cal G}_1^{-1} \equiv
{\cal G}_1^{-1}({\bf r}-{\bf r}_0,t)$. Operating with ${\cal G}^{-1}$ we may
rewrite the Wyld equation (\ref{wyld}) as
\begin{eqnarray}
\label{dyson_inv}
	& & \left[ \partial_t + \hat{\cal D}_1({\bf r}_1-{\bf r}_0) \right]
	\left[- \partial_t + \hat{\cal D}_1({\bf r}_2-{\bf r}_0) \right]
	{\cal F}({\bf r}_0|{\bf r}_1,{\bf r}_2,t)  \nonumber \\
	& & =\delta(t)
	\left[\Phi({\bf r}_0|{\bf r}_1,{\bf r}_2) +
	\Phi_0({\bf r}_0|{\bf r}_1,{\bf r}_2)
	\right],
\end{eqnarray}
where we have used the fact that ${\cal F}({\bf
r}_0|{\bf r}_1,{\bf r}_2,t)$ is only a  function of
the time difference $t = t_1 - t_2$. We have also the
``boundary" condition  \begin{equation}
\label{f_bc}
	{\cal F}({\bf r}_0|{\bf r}_1,{\bf r}_2,t) \rightarrow
	0  \hspace{0.5cm} \mbox{for} \hspace{0.5cm}
        t \rightarrow \pm \infty\ .
\end{equation}
Next we want to derive the differential equation satisfied by
the   simultaneous 2 point correlator ${\cal F}({\bf R})$:
\begin{equation}
{\cal F}({\bf R})={\cal F} ({\bf r}_1,{\bf r}_2,t=0)
\label{sim-F}
\end{equation}
where ${\bf R} = {\bf r}_1 - {\bf r}_2$. The derivation is described in
detail in Appendix A with the final result
\begin{equation}
	\hat{\cal D}_2({\bf R} ) {\cal F}({\bf R} )  = \Phi_0({\bf R} ),
\label{difference}
\end{equation}
where generally the operator $\hat{\cal D}_2({\bf r}_\alpha,{\bf r}_\beta)$
operates on
two coordinates:
\begin{equation}
\hat{\cal D}_2({\bf r}_\alpha,{\bf r}_\beta) = -\kappa [\nabla^2 _\alpha
+\nabla^2 _\beta]+ \hat{\cal B}
({\bf r}_\alpha,{\bf r}_\beta) \ , \label{D2}
\end{equation}
but in the case where the operand is a function only of the difference
${\bf R} = {\bf r}_{\alpha}-r_{\beta}$ reduces to
\begin{equation}
	 \hat{\cal D}_2({\bf R} )   \equiv
      -2\kappa \nabla^2   + {\hat{\cal B}}({\bf R} )    \ .
\label{def-D_2}
\end{equation}

\subsection{The derivation of the differential equations for higher order
correlations and response functions}

In this subsection we present the equations for the 4-point
and higher order correlation functions and for the
nonlinear Green's function (\ref{g_def}). The solutions are
deferred to the next section.

\subsubsection{The 4-point Green's function}

In paper I we presented the diagrammatic series for the
nonlinear Green's function ${\cal G}_2({\bf
r}_0|x_1,x_2,x_3,x_4)$.
This quantity is defines as
\begin{equation}
{\cal G}_2({\bf r}_0|x_1,x_2,x_3,x_4)  \equiv
	\left\langle  \frac{\delta T({x_0|x_1)}}
	{\delta \xi(x_0|x_3)}
	\frac{\delta T(x_0|x_2)}
	{\delta \xi(x_0|x_4)} \right\rangle
       	\Bigg|_{\xi \rightarrow 0} \ .
\end{equation}

The diagrammatic expansion of this quantity is an
infinite series of ladder diagrams that can be
resummed exactly. In Fig. 1 we recall the notation for
the diagrammatic elements, and in Fig. 2 is reproduced
the diagrammatic resummed equation for this function.
In analytic form this equation reads
\begin{eqnarray}
      &&{\cal G}_2(0|x_1,x_2,x_3,x_4)
={\cal G}^{^{\rm G}}_2(0|x_1,x_2,x_3,x_4)
\nonumber \\
&+&\int d{\bf r}'d{\bf r}''\int_{t_m}^\infty dt' \,
{\cal G}^{^{\rm G}}_2(0|x_1,x_2,{\bf r}',t',{\bf r}'',t')
\label{3.1}\\
	& \times& H_{ij}({\bf r}',{\bf r}'')
\frac{\partial}{\partial r'_i}
  	\frac{\partial}{\partial r_j''}
  	G_2(0|{\bf r}',t',{\bf r}'',t',x_3,x_4)
\nonumber
\end{eqnarray}
  where $t_m = \mbox{min}\{t_1,t_2\}$ and
\begin{equation}
  	{\cal G}^{^{\rm G}}_2(0|x_1,x_2,x_3,x_4) \equiv
  	{\cal G}(0|x_1,x_3){\cal G}(0|x_2,x_4).
\label{3.2}
\end{equation}
We want now to derive a differential equation for this
quantity. We can  proceed as in the case of the
correlation function by applying the product  of the
inverse operators $\tilde{{\cal G}}_1^{-1}$ of Eq.
(\ref{g_inv}) to obtain  a differential equation for
${\cal G}_2(0|x_1,x_2,x_3,x_4)$ as a function  of
three time differences. We may however make use of the
structure of Eq. (\ref{3.1}) to directly derive an
equation in only one time difference.  The
availability of such a reduction follows from the
delta-correlated  velocity field in much the same way
as the availability of a differential  operator for
${\cal F}(0|{\bf r}_1,{\bf r}_2,t=0)$ as we saw
before.  We can choose at will to consider Eq.
(\ref{3.1}) for the times $t_1=t_2 =t$  and
$t_3=t_4=0$ to derive a differential equation in one
time difference  for the quantity
\begin{equation}
	{\cal G}_2(0|{\bf r}_1,{\bf r}_2,{\bf r}_3,{\bf r}_4,t)
	\equiv {\cal G}_2
	(0|{\bf r}_1,t,{\bf r}_2,t,{\bf r}_3,0,{\bf r}_4,0)
\label{3.3}
\end{equation}
The derivation resembles closely the derivation of the
equation for the two-point correlator described in
Appendix A.  Applying the operator (\ref{dyson_inv})
to Eq. (\ref{3.1}) with the above choice of times we
get
\begin{eqnarray}
      &&[\partial_t + \hat{\cal D} _2({\bf r}_1,{\bf r}_2) +
      \hat{{\cal H}}({\bf r}_1,{\bf r}_2)]
	{\cal G}_2(0|{\bf r}_1,{\bf r}_2,{\bf r}_3,{\bf r}_4,t)
\label{3.4}  \\
&=&\delta({\bf r}_1-{\bf r}_3) \delta({\bf r}_2-{\bf r}_4)\delta(t)\,,
\nonumber
\end{eqnarray}
where $\hat{\cal D} _2({\bf r}_1,{\bf r}_2)$ is the operator defined in
Eq. (\ref{def-D_2}) and $\hat{{\cal H}}$ is an operator defined in
Appendix A that arises from the derivation of the equation for the
two-point correlator.

For the time integrated 4-point Green's function
\begin{equation}
	{\cal G}_2({\bf r}_1,{\bf r}_2,{\bf r}_3,{\bf r}_4)= \int dt
	{\cal G}_2({\bf r}_1,{\bf r}_2,{\bf r}_3,{\bf r}_4,t)
\label{Gt2}
\end{equation}
one has the equation
\begin{equation}
	\left[\hat{\cal D} _2({\bf r}_1,{\bf r}_2) +
	\hat{{\cal H}}({\bf r}_1,{\bf r}_2) \right]
	{\cal G}_2({\bf r}_1,{\bf r}_2,{\bf r}_3,{\bf r}_4)
	=\delta({\bf r}_1-{\bf r}_3) \delta({\bf r}_2-{\bf r}_4) \ .
\label{EqG2t}
\end{equation}

The non-linear Greens' function is the kernel of the
response to some forcing at points ${\bf r}_3$ and ${\bf r}
_4$, and we are interested in this response. In order to
study this, we introduce a new function
\begin{equation}
	\Psi ({\bf r}_1,{\bf r}_2) = \int d{\bf r}_3d{\bf r}_4 {\cal G} ({\bf r}
_1,{\bf r} _2,{\bf r} _3,{\bf r} _4) A({\bf r}_3,{\bf r}_4) \ ,
\label{defpsi}
\end{equation}
where $A ({\bf r}_3,{\bf r}_4)$ is an arbitrary function. From Eq.(\ref{EqG2t})
one may determine that
\begin{equation}
	\left[\hat{\cal D} _2({\bf r}_1,{\bf r}_2) + \hat{{\cal H}} \right]
	\Psi({\bf r}_1,{\bf r}_2) = A({\bf r}_1, {\bf r}_2).
\end{equation}
If one chooses now to restrict $A({\bf r}_1, {\bf r}_2)$ to the space of
functions
depending only upon the difference ${\bf R} = {\bf r}_1 - {\bf r}_2$, the
equation
simplifies (as in the derivation in Appendix A) to the form
\begin{equation}
\label{eqpsik}
	\hat{\cal D}_2({\bf R}) \Psi({\bf R}) = A({\bf R}).
\end{equation}
We will return to an analysis of this equation in Sec.3C.

\subsubsection{The simultaneous higher order correlations}
The equations of motion for the time dependent $2n$-th order
correlation functions are derived in Appendix B. Here we will
derive the equations for the simultaneous correlations.
The simultaneous $2n$-point correlator is
\begin{equation}
	{\cal F}_{2n}({\bf r}_1,{\bf r}_2,...,{\bf r}_{2n}) =
	\left< T(0|{\bf r}_1,t)T(0|{\bf r}_2,t),...,T(0|{\bf r}_{2n},t)
	\right>  .
\end{equation}
This same time quantity is identical in the Eulerian frame of
reference  and in the transformed reference frame that we use.
We can thus forget the  ${\bf r}_0$ designation, and remember
that in homogeneous systems the  quantity is a function only  of
differences of its space arguments. Its  time derivative is on
the one hand zero and on the other hand
\begin{eqnarray}
&&  	\frac{\partial}{\partial t}
  	{\cal F}_{2n}({\bf r}_1,{\bf r}_2,...,{\bf r}_{2n})
\label{3.15}\\
&=&\sum_{\alpha=1}^{2n} \left< T(0|{\bf r}_1,t)\ \dots \
	\frac{\partial}{\partial t}
  	T(0|{\bf r}_\alpha,t)\ \dots \  T(0|{\bf r}_{2n},t)\right> \ .
\nonumber
\end{eqnarray}
Using the equation of motion (\ref{advect_ql}) we find that the
RHS of (\ref{3.15})  has three types of terms, one with ${\bf
w}$ (advection), one with  $\kappa$ (dissipation) and the last
proportional to the force (forcing).   These terms are denoted
as $A_{\rm adv}$, $A_{\rm dis}$ and $A_{\rm for}$  with
\begin{equation}
	A_{\rm adv} + A_{\rm dis} + A_{\rm for} = 0\,
\label{3.16}
\end{equation}
where
\begin{equation}
	A_{\rm adv} = \sum_{\alpha=1}^{2n} \nabla_\alpha \cdot
	{\bf F}_{w,2nT}({\bf r}_\alpha,{\bf r}_1 ...{\bf r}_{2n})
\end{equation}
and
\begin{eqnarray}
      &&{\bf F}_{w,2nT}({\bf r}_\alpha,{\bf r}_1\ \dots \ {\bf r}_{2n})
\label{3.18}\\
 & = & \left< {\bf w }_\alpha T(0|{\bf r}_1,t)\dots T(0|{\bf
   r}_{\alpha-1},t)T(0|{\bf r}_\alpha,t)\dots  T(0|{\bf r}_{2n},t)
	\right>
\nonumber \\
&& A_{\rm dis}  =  \kappa \sum_{\alpha=1}^{2n}   \nabla^2_\alpha
	{\cal F}_{2n}({\bf r}_1,{\bf r}_2,\ \dots \ {\bf r}_{2n})\ .
\label{3.19}
\end{eqnarray}
Finally
\begin{equation}
\!  	A_{\rm for}\! =\! \sum_{\alpha=1}^{2n}  \left<
  	T(0|{\bf r}_1\!, t)... T(0|{\bf r}_{\alpha-1}, t)
  	\phi(0|{\bf r}_\alpha\!,t)...
  	T(0|{\bf r}_{2n}\!,t)\right>.
\label{3.20}
\end{equation}
The diagrammatic representation of ${\bf F}_{w,2nT}({\bf
r}_\alpha,{\bf  r}_1...{\bf r}_{2n})$ was discussed in paper I
with a final result [cf.  paper I Eq. (5.7)] which in ${\bf
r},t$ representation  is
\begin{eqnarray}
     &&	{\bf F}_{w,2nT}({\bf r}_\alpha,{\bf r}_1...{\bf r}_{2n})
\label{3.21} \\
       &=&  \frac{1}{2} \sum_{\beta=1}^{2n}
  	{\bf H}(0|{\bf r}_\alpha,{\bf r}_\beta)\cdot\nabla_\beta
	{\cal F}_{2n}({\bf r}_1,{\bf r}_2,...,{\bf r}_{2n})
\nonumber
\end{eqnarray}
Introducing this result into (\ref{3.18}) and using (\ref{h_euler})
we find
\begin{eqnarray}
 	A_{\rm adv} &=&   \frac{1}{2}\sum_{\alpha,\beta=1}^{2n}
  	\left[h_{ij}({\bf r}_\alpha)+h_{ij}({\bf r}_\beta)-
  	h_{ij}({\bf r}_\alpha-{\bf r}_\beta)\right]
 \nonumber   \\
&\times&\frac{\partial^2}{\partial r_{\alpha i} \partial r_{\beta j}}
  	{\cal F}_{2n}({\bf r}_1,{\bf r}_2,...,{\bf r}_{2n}).
\label{3.22}
\end{eqnarray}
This form can be rewritten equivalently as
\begin{eqnarray}
    &&    A_{\rm adv} = \Bigg\{ -  \sum_{\alpha> \beta=1}^{2n}
\hat{\cal B} _{\alpha\beta}
\label{3.23}  \\
&&+\Bigg[ \sum_{\alpha=1}^{2n}
  	 h_{ij}({\bf r}_\alpha)
  	\frac{\partial}{\partial r_{\alpha i}} \Bigg]
  	 \Bigg[\sum_{\beta=1}^{2n} \frac{\partial}{\partial r_{\beta j}}
         \Bigg]
      \Bigg\}
{\cal F}_{2n}({\bf r}_1,{\bf r}_2,...,{\bf r}_{2n}). \nonumber
\end{eqnarray}
Finally we use the fact that ${\cal F}_{2n}$ is a function of
differences of its spatial arguments to see that the second
operator in the first line of the RHS gives zero and the
line can be omitted. Thus
\begin{equation}
        A_{\rm adv} =
 -  \sum_{\alpha> \beta=1}^{2n}
\hat{\cal B} _{\alpha\beta}
	{\cal F}_{2n}({\bf r}_1,{\bf r}_2,...,{\bf r}_{2n}) .
\label{3.24}
\end{equation}
Consider next the forcing term. Using the fact that for a
Gaussian force
\begin{eqnarray}
	& & \left< T(0|{\bf r}_1,t)...T(0|{\bf r}_{\alpha-1},t)
	\phi(0|{\bf r}_\alpha,t)...T(0|{\bf r}_{2n},t)\right>
\nonumber \\
&=&\int     \left< \frac{\delta T(0|{\bf r}_1,t)...T(0|{\bf r}_{\alpha-1},t)
 	...T(0|{\bf r}_{2n},t)}{\delta\phi(0|{\bf r}_\alpha',t')} \right>
\nonumber \\
&\times& \Phi_0({\bf r}_\alpha-{\bf r}_\alpha') d{\bf r}_\alpha'
\label{3.25}
\end{eqnarray}
where we used (1.2) and the fact that we deal with simultaneous
correlations. The functional derivative in the integrand is a
zero time  response, which as usual  is computed in the
non-interacting limit:
\begin{eqnarray}
&&  	\left<
  	\frac{\delta T(0|{\bf r}_1,t)...T(0|{\bf r}_{\alpha-1},t)
  	...T(0|{\bf r}_{2n},t)}{\delta\phi(0|{\bf r}_\alpha',t')}
  	\right>
\label{3.26}  \\
       &=& \sum_\beta G^0({\bf r}_\alpha-{\bf r}_\beta,t=0)
  	{\cal F}_{2n-2}({\bf r}_1,{\bf r}_2,...{\bf r}_{2n})
\nonumber
\end{eqnarray}
where in ${\cal F}_{2n-2}({\bf r}_1,{\bf r}_2,...{\bf
r}_{2n})$ the two  arguments ${\bf r}_\alpha$ and ${\bf
r}_\beta$ are missing. Substituting (\ref{3.26}) in (\ref{3.25})
and the result in (\ref{3.20}), we find
  \begin{equation}
  A_{\rm for}= \sum_{\alpha>\beta} \Phi_0({\bf r}_\alpha-{\bf
  r}_\beta) {\cal F}_{2n-2}({\bf r}_1,{\bf r}_2,...{\bf r}_{2n}).
\label{3.27}
   \end{equation}
Substituting all these results in (\ref{3.16}) yields
\begin{eqnarray}
        & & \left[- \kappa \sum_{\alpha} \nabla^2_\alpha +
  	\sum_{\alpha>\beta}^{2n}
        \hat{\cal B} _{\alpha\beta}
        \right]
  	{\cal F}_{2n}({\bf r}_1,{\bf r}_2,...,{\bf r}_{2n})
  \nonumber \\
   &	&=
	\sum_{\alpha>\beta}
	\Phi_0({\bf r}_\alpha-{\bf r}_\beta)
	{\cal F}_{2n-2}({\bf r}_1,{\bf r}_2,...{\bf r}_{2n})\ .
\label{3.28}
\end{eqnarray}
This equation, which is our final equation for the
simultaneous correlation function, is again identical with
Kraichnan's.

\subsubsection{Equation for the irreducible correlation
${\cal F} _4^{\rm c}$}

The irreducible (cumulant) part of the correlation  functions
are defined as
\begin{equation}
        {\cal F}_{2n}^{\rm c}={\cal F}_{2n}-{\cal F}_{2n}^G.
\end{equation}
The equation for the simultaneous ${\cal F}_4^{\rm c}$ follows from
specializing Eq. (\ref{3.28})  to the case $2n=4$, using Eq.
(\ref{difference}) for ${\cal F} _2$,
\begin{eqnarray}
\label{cumF}
        & & \left[- \sum_{\alpha}^4 \kappa\nabla^2_\alpha +
\sum_{\alpha>\beta} \hat{\cal B} _{\alpha\beta} \right]
\sum_{\alpha>\beta=1}^4
        {\cal F}_{4}^{\rm c}({\bf r}_1,{\bf r}_2,{\bf r}_3,{\bf r}_4)
  \nonumber \\
   &    &=
     -\frac{1}{2}\sum_{\scriptstyle(\alpha_i)\atop{\rm perm} (1234)}
   \hat{\cal B} ({\bf r}_{\alpha_1}-{\bf r}_{\alpha_3})
  {\cal F}_2({\bf r}_1,{\bf r}_2){\cal F}_2({\bf r}_3,{\bf r}_4) \\
& & =-\Big[(\hat{\cal B} _{14}+ \hat{\cal B} _{23}+\hat{\cal B} _{13}+\hat{\cal
B} _{24}){\cal F} _{12} {\cal F} _{34}
\nonumber\\
&+&(\hat{\cal B} _{14}+ \hat{\cal B} _{23}+\hat{\cal B} _{12}+\hat{\cal B}
_{34}){\cal F} _{13} {\cal F} _{24}
\nonumber\\
&+& (\hat{\cal B} _{13}+ \hat{\cal B} _{24}+\hat{\cal B} _{12}+\hat{\cal B}
_{34}){\cal F} _{14} {\cal F} _{23}
     \Big] \ .
\nonumber
\end{eqnarray}

\section{Analysis of the Green's functions and the 2-point
correlation}
In this section we begin to discuss the exact solution for the
various quantities in  this model when all the separation
distances are in the inertial interval. For the sake of clarity
we catalogue here all the equations that we are  going to solve
below, neglecting the diffusive terms which appear in the full
equations.  The equations are the following, for

1) the time integrated Green's function cf.(\ref{Gt}):
\begin{equation}
	\hat{\cal B} ({\bf R} ) {\cal G}({\bf R} ,{\bf R}') =
	\delta ({\bf R} - {\bf R}') \ , \label{EqG}
\end{equation}
which follows from (\ref{EqGt}),

2) the 2 point simultaneous correlation functions,
\begin{equation}
\hat{\cal B} ({\bf R} ) {\cal F}({\bf R} ) = \Phi ({\bf R} ) \,,\label{EqF2}
\end{equation}
which follows from (\ref{difference}),

3) the function $\Psi$ introduced in Sec.2B,
\begin{equation}
	\hat{\cal B} ({\bf R}) \Psi({\bf R}) = A({\bf R}),
\label{eqpsi}
\end{equation}
which is the $\kappa \to 0$ limit of Eq.(\ref{eqpsik}),

4) the cumulant of the 4-point simultaneous correlation function
\begin{eqnarray}
	&&\Big[\hat{\cal B} _{12}+\hat{\cal B} _{13}+\hat{\cal B} _{14}+\hat{\cal B}
_{23}+
	\hat{\cal B} _{24}+\hat{\cal B} _{34}\Big]{\cal F}^c({\bf r}_1,{\bf r}_2,{\bf
r}_3,{\bf r}_4)
\nonumber\\
	&=& -\Big[\hat{\cal B} _{14}+\hat{\cal B} _{23}+\hat{\cal B} _{13}+\hat{\cal
B} _{24}\Big]
	{\cal F}({\bf r}_1-{\bf r}_2){\cal F}({\bf r}_3-{\bf r}_4)
\nonumber\\&-&\Big[\hat{\cal B} _{14}+\hat{\cal B} _{23}+\hat{\cal B}
_{12}+\hat{\cal B} _{34}\Big]
	{\cal F}({\bf r}_1-{\bf r}_3){\cal F}({\bf r}_2-{\bf r}_4)
\nonumber\\&-&\Big[\hat{\cal B} _{13}+\hat{\cal B} _{24}+\hat{\cal B}
_{12}+\hat{\cal B} _{34}\Big]
	{\cal F}({\bf r}_1-{\bf r}_4){\cal F}({\bf r}_2-{\bf r}_3) \ ,
\label{EqF4}
\end{eqnarray}
which is the $\kappa\to 0$ limit of eq.(\ref{cumF}).

It is noteworthy that the same operator $\hat{\cal B}$, (\ref{def-B2}),
(\ref{sep-B})  appears in all these equations. Indeed, the  most
interesting scaling properties will be seen to arise from the
eigenfunctions of $\hat{\cal B}$  with zero eigenvalue which are the
solutions of the homogeneous equation (\ref{EqGhom}).

\subsection{Solutions of the basic homogeneous equation}

For two distinct coordinates ${\bf R} \ne {\bf R} '$ the Green's function
${\cal G}({\bf R} ,{\bf R}')$ satisfies the homogeneous part of the equation
(\ref{EqG})
\begin{equation}
	\hat{\cal B} ({\bf R} ) {\cal G}({\bf R} ,{\bf R}') = 0   \ , \label{EqGhom}
\end{equation}
which will be referred to hereafter as the `` basic homogeneous
equation". In light of the representation  of Eq.(\ref{sep-B})
of the $\hat{\cal B}$ operator via the angular momentum operator we
represent the solution of  (\ref{EqGhom}) in 3-dimensional
space as an expansion over spherical harmonics
\begin{equation}
	{\cal G}({\bf R} ,{\bf R}') =
\sum_{l=0}^{\infty}\sum_{m=-l}^l\sum_{l'=0}^{\infty}
	\sum_{m'=-l'}^{l}b_{ll',mm'}
	g_{lm}({\bf R} )g_{l'm'}({\bf R}' )
\label{sumharm}
\end{equation}
where $b_{ll',mm'}$ are coefficients and
\begin{equation}
	g_{lm}({\bf R} ) = R^{\beta_l}Y_{lm}(\theta,\phi) \ .
\label {harm}
\end{equation}
The spherical harmonics $Y_{lm}(\theta,\phi)$ are the
eigenfunctions of the  angular momentum operator:
\begin{equation}
	\hat {L}^2Y_{lm}(\theta,\phi)=l(l+1)Y_{lm}(\theta,\phi)
\end{equation}
and $\theta$ and $\phi$ are the polar and azimuthal angles of
${\bf R}$. Substituting the expansion (\ref{sumharm})into
eq.(\ref{EqGhom}) we find the relationship
\begin{equation}
	\beta_l(\beta_l+3-\zeta_2)-l(l+1)(1+\zeta_h/2)=0  \ .
\label{relate}
\end{equation}
Of the two solutions of this quadratic equation we must select
the nonnegative branch since the negative branch is unphysical:
\begin{equation}
	\beta_l={1 \over 2}\Bigg[\zeta_2-3 +\sqrt{(\zeta_2-3)^2 +
	2l(l+1)(2+\zeta_h)}\Bigg] \ . \label{betal3}
\end{equation}
In 2 dimensions one has a similar representation with the
difference that instead of the spherical harmonics we have the
eigenfunctions $\exp(il\phi)$ and instead of the eigenvalues
$l(l+1)$ we have the eigenvalues $l^2$. The final result for the
exponents $\beta_l$ in 2- dimensions is
\begin{equation}
	\beta_l={1 \over 2}\Bigg[-\zeta_h+
	\sqrt{\zeta_h^2 +4l^2(1+\zeta_h)}\Bigg] \ . \label{betal2}
\end{equation}

Multipole expansions of the type (\ref{sumharm}) and the scaling
exponents $\beta_l$ play an important role in all our
development below. We will see that the basic homogeneous
equation reappears in various guises, and in each of them we
will need to fix the coefficients $b_{ll',mm'}$ such as to
respect the symmetry of the particular object involved, the
boundary conditions, etc. For example in this case we know that
the Green's function ${\cal G} ({\bf R} ,{\bf R}')$ is symmetric in ${\bf R}$
and
${\bf R}'$. Accordingly $b_{ll',mm'} = b_{l'l,m'm}$. If the solution
depends on $R,\ R'$ and the angle between ${\bf R}$ and ${\bf R}'$, then
$b_{ll',mm'}\neq 0$ only for $l=l',\ m+m'=0$, etc. 

\subsection{Exact solution of the 2-point correlation function}
\subsubsection {Isotropic forcing}
The solution of the 2-point correlation function depends on
the nature of the external forcing $f({\bf r},t)$. It is
customary to take $f({\bf r},t)$ to be Gaussian and
statistically homogeneous in space and time. The properties of
the correlation function $\Phi_0({\bf r}-{\bf r}')$ were not
determined up to now. Since we are interested in the universal
scaling properties of the scalar field we want to choose the
forcing such that it has only large scale components.
Otherwise the scaling exponent of the 2-point correlation may be
coloured by the functional dependence of the forcing on $r$. On
the other hand the forcing may be isotropic or non isotropic. In
this subsection we will  deal with the isotropic case, and the
nonistropic case  will be treated in the next subsection.

The properties of $f({\bf r},t)$ are best stated in ${\bf
k}$-space:  it is concentrated in the small $k$ region, i.e.
$k \leq 1/L$, and it decays quickly to zero for $k \ll 1/L$.
In ${\bf r}$ space this means that $\langle f({\bf r},t)
f({\bf r}+{\bf R},t) \rangle$ is constant for ${\bf R}\ll L$:
\begin{equation}
	\langle f({\bf r},t) f({\bf r}+{\bf R},t) \rangle =
	\Phi_0 .  \hspace{0.5cm} (R \ll L)
\end{equation}
Using this form of the forcing correlation function we realize
that the inhomogeneous term on the RHS of Eq.(\ref{sim-F})
is translationally and rotationally invariant. Accordingly
we can seek solutions that have the same symmetry, solving the equation
\begin{equation}
	\hat B(R) {\cal F} (R) = \Phi_0  \ .\label{iso}
\end{equation}

The solution of the inhomogeneous equation is
\begin{equation}
	{\cal F}_{\rm inh}(R) =C_{\rm inh}R^{\zeta_2} \ , \label{Fin}
\end{equation}
with $C_{\rm inh} = - \Phi_0/2h \zeta_2$ and
\begin{equation}
	\zeta_2 = 2-\zeta_h \ . \label{zeta2}
\end{equation}
Of course, the inhomogeneous solution has to be supplemented
with  the solutions of the homogeneous equation in order to
match the boundary conditions (which in this case are ${\cal F} (R) =
{\cal F} (0)\neq 0$  at $r=0$). There are two homogeneous solutions:
one is a constant which  must be taken as ${\cal F} (0)$, and the
other is  ${\cal F}(R) \propto R^{-\zeta_h}$ which is ruled out
by the boundary condition ${\cal F}(0) < \infty$. One should
note that the constant homogeneous solution belongs to the
family of exponents $\beta_l$ with $l=0$ as one would expect.
The solution that is ruled out is indeed the $l=0$ member of the
forbidden branch of solutions of the quadratic equation
(\ref{relate}).
\subsubsection{The law of isotropization}

When the forcing is nonisotropic on the large scales we need to
apply the full operator $\hat{\cal B}({\bf R} )$. In the regime $R\ll L$ the
forcing is again a constant which is independent of the angles.
However, for $R\simeq L$ we expect nonisotropic forcing  which
leads to nonisotropic tails for every value of R. We can find
the law of isotropization of ${\cal F} ({\bf R} )$ for $R$ small by
solving the equation that involves now also the nonisotropic
part of the operator $\hat{\cal B} ({\bf R})$
\begin{equation}
	\hat{\cal B}({\bf R}) {\cal F} ({\bf R}) =
	\Phi_0 \ , \quad  R\ll L .\label{noniso}
\end{equation}
The inhomogeneous solutions are the same as before,
but in order to match with the anisotropic behaviour
at $R\simeq L$ we need to invoke the whole set of
homogeneous solutions $g_{lm}({\bf R} )$, Eq. (\ref{harm}).
The solution in 3 dimensions is written as
\begin{equation}
	{\cal F} ({\bf R}) = {\cal F}(0) +
	R^{\zeta_2}\Bigg[C_{in}+
	\sum_{l=2}^{\infty}\sum_{m=-l}^l a_{lm}
	\left({R \over L}\right)^{\beta_l-\zeta_2} Y_{lm}
	(\theta,\phi)\Bigg] .
\label{lawiso}
\end{equation}
To understand this result we note that all values of $\beta_l$
are larger than $\zeta_2$. In particular $\beta_2$ (for both
$d=3$ and $d=2$) is
\begin{equation}
	\beta_2 = {1 \over 2}\Bigg[\zeta_2-d + \sqrt{(\zeta_2+d)^2 +
	24\zeta_h}\Bigg] \ . \label{beta2}
\end{equation}
For positive $\zeta_h$, $\beta_2$ is larger than $\zeta_2$. For
larger $l$, $\beta_l$ is even larger, and dependent on dimension.
Thus all the anisotropic
terms decay when $R\ll L$. Note that  the coefficients $a_{lm}$
are nonuniversal and should be found by matching at $R\sim L$.
The law of decay is however universal.
\subsubsection{Anisotropic structure functions}

The solution (\ref{lawiso}) suggests the introduction of
anisotropic structure functions  via the definition
\begin{equation}
	S_{2,lm}(R) =\int d\theta d\phi Y_{lm}(\theta,\phi) S_2({\bf R}) \ .
\label{S2an}
\end{equation}
Here $l$ should be even due to the symmetry with respect to the
inversion of ${\bf R}$. These anisotropic structure functions display
"clean" scaling behavior with the exponent $\beta_l$:
\begin{equation}
	S_{2,lm}(R) \sim S_2(L) \left( {R\over L}\right)^{\beta_l} \ .
\label{S2scale}
\end{equation}
We will see that the same scaling exponents feature prominently below.

\subsection{The nonlinear Green's function and the anomalous
exponent $\Delta$}

In this subsection we discuss the nonlinear Green's function and
the anomalous exponent $\Delta$ which is associated with it, cf.
(\ref{anom}).  To this aim we return to the equation for $\Psi$,
Eq.(\ref{eqpsi}).
We realize that the solutions of this equation are identical
to those discussed already in the context of the 2-point
correlator. For a constant function $A$ we can write the
solution
\begin{equation}
	\Psi({\bf R} ,{\bf R} ') = \Psi (0) + C|{\bf R}-{\bf R}'|^{\zeta_2} \ .
\label{solpsi}
\end{equation}
Taking the derivative with respect to ${\bf R}$ and ${\bf R}'$ as required
by Eq.(\ref{anom}) we find
\begin{equation}
	(\bbox{\nabla} _1\cdot\bbox{\nabla} _2)\Psi({\bf R} ,{\bf R} ') \sim |{\bf
R}-{\bf R}'|^{-\zeta_h}
\label{diverge}
\end{equation}
Comparing now with the definition of the anomalous exponent
$\Delta$ in Eq.(\ref{anom}) we see that in the limit $|{\bf R}-{\bf R}'|\to
\eta$ this quantity diverges like $1/\eta^{\zeta_h}$ and
\begin{equation}
\Delta = \zeta_h . \label{delta}
\end{equation}
As explained before this value of $\Delta$ is the critical value $\Delta_c$,
and the implications of this finding are explored below. We
turn now to the appearance of the anomalous exponent  in the
4-point correlator and related quantities.

\section{The four point correlator and related quantities:
$K(R)$, and $J_4 (R)$ and $L(R)$}

In this section we present the analysis of the four point
correlator ${\cal F}({\bf r} _1,{\bf r} _2,{\bf r} _3,{\bf r} _4)$ and of the
quantities that are related
it, i.e the correlation of dissipation fluctuation $K(R)$,
Eq.(\ref{K}) and the RHS of the balance equation $J_4 (R)$,
Eq.(\ref{J}). These last  quantities are not exactly 4-point
correlations, but they are obtained as a limit of ${\cal F} _4({\bf r} _1,{\bf
r} _2,{\bf r} _3,{\bf r} _4)$.
The simpler limit is $K(R)$ which is a centred correlation
function and is therefore related to ${\cal F} _4 ({\bf r} _1,{\bf r} _2,{\bf
r} _3,{\bf r} _4)$:
\begin{eqnarray}
	K(R) &= &
	\kappa^2\lim_{r_{12},r_{34}\to 0}
	\lim_{r_{13}\to R}(\bbox{\nabla} _1\cdot\bbox{\nabla} _2)
	(\bbox{\nabla} _3\cdot\bbox{\nabla}_4) \nonumber \\
&\times &\Big[{\cal F}^c _4 ({\bf r} _1,{\bf r} _2,{\bf r} _3,{\bf r} _4)+
	{\cal F}_2({\bf r}_1,{\bf r}_3){\cal F}_2({\bf r}_2,{\bf r}_4)\Big] \ .
\label{limK}
\end{eqnarray}
Since we discovered that the the second derivative is singular ,
cf. Eq.(\ref{diverge}), we need to carefully examine the above
limit.

The quantity $J_4 (R)$ has a Gaussian decomposition $J_4^G (R)$
and a cumulant part $J_4^c (R)$.  The Gaussian decomposition is
trivially computed as
\begin{equation}
	J_4^G (R) = \bar\epsilon S_2(R) + {\kappa\over 2}
	[\nabla S_2(R)]^2 \ .
\end{equation}
The cumulant part is related to ${\cal F}^c _4$
via two terms, $J_4^c (R)=J_{4,1} (R) + J_{4,2} (R)$,
\begin{eqnarray}
	J_{4,1} (R)&=&\kappa
	\left< \left<|\bbox{\nabla} _1
	\Theta({\bf r}_1)|^2\Theta^2({\bf r}_2)\right>\right> \ ,
\label{J41} \\
	J_{4,2} (R)&=&-2\kappa\left<\left<|\bbox{\nabla} _1\Theta({\bf r}_1)|^2
	\Theta({\bf r}_1)\Theta({\bf r}_2)\right>\right> \ ,
\label{J42}
\end{eqnarray}
where double brackets denote the cumulant part. These two contributions
may be considered as the following limits of ${\cal F}^c({\bf r} _1,{\bf r}
_2,{\bf r} _3,{\bf r} _4)$
\begin{eqnarray}
	J_{4,1} (R)&=&-\kappa\lim_{r_{12},r_{34}\to 0}
	\lim_{r_{13}\to R}(\bbox{\nabla} _1\cdot\bbox{\nabla} _2)
	{\cal F}^c _4 ({\bf r} _1,{\bf r} _2,{\bf r} _3,{\bf r} _4) \ ,
\label{limJ1}\\
	J_{4,2} (R)&=&2\kappa\lim_{r_{12},r_{13}\to 0}
	\lim_{r_{14}\to R}(\bbox{\nabla} _1\cdot\bbox{\nabla} _2)
	{\cal F}^c _4 ({\bf r} _1,{\bf r} _2,{\bf r} _3,{\bf r} _4) \ .
\label{limJ2}
\end{eqnarray}
Note that in Eq.(\ref{limJ1}) we have two pairs of coalescing
points, i.e {1,2} and {3,4},  which are separated by a large
distance $R$. In Eq.(\ref{limJ2}) we have three coalescing
points, i.e. {1,2,3} and this group is separated from point 4 by
$R$. One should also note that $J_4$ is obtained from the full
${\cal F}_4$ and not from the cumulant part.

In the next subsections we are going to make strong use of the
divergence with respect to  small distances. Our strategy will
be to expose the leading exponent in the divergence with respect
to small separations and to compute it exactly. Then we will
find the exponent of the dependence on $R$ by power counting,
knowing the overall scaling exponent $\zeta_4$ of  ${\cal F}^c _4$.
In other words, our basic assumption is that the correlator
${\cal F}^c_4$ is a homogeneous function of its arguments as long as
all of them are in the inertial range:
\begin{equation}
	{\cal F}^c_4(\lambda{\bf r}_1,\lambda{\bf r}_2,\lambda{\bf r}_3,\lambda{\bf
r}_4)=
	\lambda^{\zeta_4}{\cal F}^c_4({\bf r} _1,{\bf r} _2,{\bf r} _3,{\bf r} _4) \ ,
\label{F4lam}
\end{equation}
where $\zeta_4$ is the unknown scaling exponent that
characterizes the structure  function $S_4(R)$. We will not make
any assumption about the numerical value of $\zeta_4$.

\subsection{Two coalescing pairs of points}\label{4A}
\subsubsection{The effective equation for ${\cal F}^c _4$}

Consider Eq.(\ref{cumF}) in the limit $r_{12},r_{34}\ll R$, but
all separations in the inertial interval. This allows us to
neglect the diffusion terms, and to conclude that $\hat{\cal B}
_{12}\propto r^{-\zeta_2}_{12}$ and $\hat{\cal B} _{34}\propto
r^{-\zeta_2}_{34}$ are much larger than all the other $\hat{\cal B}
_{\alpha\beta}$ operators which are proportional to
$R^{-\zeta_2}$. This suggests to rewrite the equation in the
form
\begin{equation}
	\Big[\hat{\cal B} _{12}+\hat{\cal B} _{34} +
	\sum\hat{\cal B}\Big]
	{\cal F}^c _4({\bf r} _1,{\bf r} _2,{\bf r} _3,{\bf r} _4) =
	{\rm RHS} \ , \label{simF4}
\end{equation}
where the sum is on all the four $\hat{\cal B}$ operators other than the
ones explicitly  displayed. The calculation of the RHS of
Eq.(\ref{cumF}) is elementary since we know everything
explicitly. The result of the calculation is
\begin{equation}
	{\rm RHS} = R^{\zeta_2}
	\Bigg[A_0+A_1 \left({r_{12}^{\zeta_h}+r_{34}^{\zeta_h}
	\over R^{\zeta_h}}\right)+
	A_2\left({r_{12} r_{34}\over R^2}\right)^{\zeta_2}\Bigg]
\ , \label{RHS}
\end{equation}
where $A_0$, $A_1$ and $A_2$ are some dimensionless constants.
The last term in the bracket comes from the first group of terms
in the RHS of Eq.(\ref{cumF}). Indeed, the product of two
${\cal F}_2$s is proportional to  $(r_{12} r_{34})^{\zeta_2}$. The
naive evaluation of the sum of $\hat{\cal B}$'s is $R^{\zeta_h}/r_{12}
r_{34}$. However there are two cancellations in the  combination
of $\hat{\cal B}$'s which result in the replacement of $r_{12} r_{34}$ by
$R^2$. The sum of $\hat{\cal B}$'s becomes proportional to
$R^{-\zeta_2}$. The leading contribution in the last two groups
of terms is regular in $r_{12}$ and $r_{34}$ and therefore can
be evaluated as $R^{\zeta_2}$. This is the first term in the
bracket. The second term in the bracket is contributed by the
 sum of $\hat{\cal B} _{12} + \hat{\cal B} _{34}$ in the second and third
groups of terms. Higher order terms exist but they are
proportional to the second power of the small  distances.

To discuss  Eq.(\ref{simF4}) further we note that in a space
homogeneous situation ${\cal F}^c _4$ is a function of six
differences in 3 dimensions and 5 in 2 dimensions. In light of
our strategy it is convenient to choose the variables  as $R$,
${\bf r}_{12}$ and ${\bf r}_{34}$. In doing so we are using seven rather
than the  needed number of variables, but we will see that the
dependence on the extra angle variables disappears. We can now group
the RHS of Eq.(\ref{simF4}) together with $\sum\hat{\cal B}{\cal F}^c _4$
into a new function, say $E({\bf r}_{12},{\bf r}_{34},R)$
\begin{equation}
E({\bf r}_{12},{\bf r}_{34},R) = {\rm RHS}-\sum\hat{\cal B}{\cal F} _4 \ .
\label{Eex}
\end{equation}
In the limits $r _{12},r _{34}\ll R$ we can
expand $E({\bf r}_{12},{\bf r}_{34},R)$ in orders of $r_{12}$ and $r_{34}$
\begin{equation}
	E({\bf r}_{12},{\bf r}_{34},R) = E^{(0)} (R) +
	E^{(1)}({\bf r}_{12},{\bf r}_{34},R) + \dots \ ,
\end{equation}
where
\begin{equation}
	 E^{(0)} (R) = \lim_{r _{12},r _{34}\to 0}
	 E({\bf r}_{12},{\bf r}_{34},R) \ .
\label{ER}
\end{equation}
The  equation that we need to analyze takes on the form
\begin{equation}
	\Big[\hat{\cal B} _{12}+\hat{\cal B} _{34}\Big]{\cal F}^c _4({\bf r}_{12},{\bf
r}_{34},R) =
	E^{(0)} (R) +
 	E^{(1)}({\bf r}_{12},{\bf r}_{34},R) + \dots \  \ .
\label{redF4}
\end{equation}
and the explicit expression for  $E^{(1)}({\bf r}_{12},{\bf r}_{34},R)$
will be presented below.
\subsubsection{ Solutions}\label{solutions}
To leading order we have the effective equation:
\begin{equation}
	\Big[\hat{\cal B} _{12}+\hat{\cal B} _{34}\Big]{\cal F}^c _4({\bf r}_{12},{\bf
r}_{34},R) =
	E^{(0)}(R) \ . \label{finalF4}
\end{equation}
The leading inhomogeneous solution of Eq.(\ref{finalF4}) is
found by inspection:
\begin{equation}
	{\cal F} _{4,\rm inh}^{(1)}({\bf r}_{12},{\bf r}_{34},R) =
	C_1 E^{(0)}(R)(r _{12}^{\zeta_2} +r _{34}^{\zeta_2}) \ ,
\label{inhF4}
\end{equation}
where $C_1$ is a dimensionless constant. Using the overall
scaling exponent $\zeta_4$ that was  introduced in
Eq.(\ref{F4lam}) we can rewrite this solution as
\begin{equation}
	{\cal F} _{4,\rm inh}^{(1)}({\bf r}_{12},{\bf r}_{34},R) \sim S_4(R)
	{{ r_{12}^{\zeta_2} +r _{34}^{\zeta_2}} \over R^{\zeta_2}} \ .
\label{inhF4f}
\end{equation}
This inhomogeneous solution will be particularly important in
the context of the calculation of $J_4(R)$. It does not
contribute however to the evaluation of $K(R)$,  and for the the
latter we need to find the next order inhomogeneous solution.
The next order term on the RHS of Eq.(\ref{redF4}) stems from
two sources. First is the $A_1$ term in Eq.(\ref{RHS}), and the
second is found by substituting (\ref{inhF4f}) into  $\sum \hat{\cal B}
{\cal F}^c _4$ on the RHS of (\ref{ER}). Both contributions have the
same dependence on $r_{12}$ and $r_{34}$, and they can be
written as:
\begin{equation}
	E^{(1)}(r_{12},r_{34},R) =
	E_1(R){{ r_{12}^{\zeta_2} +r _{34}^{\zeta_2}} \over R^{\zeta_2}} \ .
\label{E1}
\end{equation}
Substituting this in Eq.(\ref{redF4}) produces the next order
inhomogeneous solution which reads
\begin{equation}
	{\cal F} _{4,\rm inh}^{(2)}({\bf r}_{12},{\bf r}_{34},R) =
	C_1 E^{(1)}(R) \left({r _{12} r _{34}\over
R}\right)^{\zeta_2}
\label{inh2F4}
\end{equation}
We can again use Eq.(\ref{F4lam}) to rewrite this solution in the form
\begin{equation}
	{\cal F} _{4,\rm inh}^{(2)}({\bf r}_{12},{\bf r}_{34},R) \sim
	S_4(R) \left({r _{12} r _{34}\over
	R^2}\right)^{\zeta_2} \ . \label{inhF4f2}
\end{equation}
This solution
will be shown to give the leading order contribution to $K(R)$,
and therefore these orders of the inhomogeneous solution will
suffice for our analysis.

In addition to the  inhomogeneous solutions all the homogeneous
solutions (\ref{sumharm}) found below are available to us, since
the homogeneous solutions of $\hat{\cal B} _{12}$ and of  $\hat{\cal B} _{34}$
are
also homogeneous solutions of $\hat{\cal B} _{12} + \hat{\cal B} _{34}$. We can
therefore write the homogeneous solution as
\begin{equation}
	{\cal F} _{4,\rm hom}({\bf r}_{12},{\bf r}_{34},R)
=\sum_{l=0}^{\infty}\sum_{m=-l}^l
	\sum_{l'=0}^{\infty}\sum_{m'=-l'}^{l}f_{ll',mm'}(R)
	g_{lm}({\bf r}_{12} )g_{l'm'}({\bf r}_{34} )\ . \label{Fsumharm}
\end{equation}
One should note that at this point we can make a choice of the
coordinate system such that the $z$ axis is directed along the
${\bf R}$ axis. Accordingly the dependence on  the sum of azimuthal
angles $\phi_{12}$ and $\phi_{34}$ should disappear due to the
symmetry of the problem. This requirement is met if all
coefficients $f_{ll',mm'}$ vanish when $m+m'\neq 0$. In addition
our correlation function is symmetric with respect to the
exchange of any pair of points. Accordingly all odd values of
$l$ and $l'$ are excluded from the sums in (\ref{Fsumharm}).
Finally we can find the $R$ dependence of these coefficients
from the overall power counting. Since $g_{l,m}({\bf r}) \propto
r^{\beta_l}$ we can write
\begin{equation}
f_{ll',mm'}(R) \sim S_4(R) /R^{\beta_l+\beta_l'}  \ . \label{fllmm}
\end{equation}

We will need below also the next order inhomogeneous solution
which is obtained by substituting (\ref{Fsumharm}) back into the
definition of $E$ (\ref{Eex}). This solution will be denoted as
\begin{equation}
	{\cal F} _{4,\rm inh}^{(3)}({\bf r}_{12},{\bf r}_{34},R) \sim
	{\cal F} _{4,\rm hom}^{(2)}({\bf r}_{12},{\bf r}_{34},R)
	{r_{12}^{\zeta_2}+r_{34}^{\zeta_2}\over R^{\zeta_2}} \ .
\label{F4(3)}
\end{equation}
In addition to these homogeneous solutions one can consider also
the homogeneous solutions of the sum of the two $\hat{\cal B}$ operators
on the LHS of (\ref{finalF4}). It can be shown that these
homogeneous solutions do not add any information that is not
contained in the solutions described above.
\subsection{Three coalescing points}

For the calculation of $J_4$ we need also to consider the
geometry of three coalescing points, see Eq.(\ref{J42}).
Accordingly we focus on the limit $r_ {12},r_ {13},r_ {23}\ll
R\simeq r_ {14} \simeq r_ {24}\simeq r_ {34}$. Instead of
Eq.(\ref{redF4}) we now have the equation
\begin{equation}
	\Big[\hat{\cal B} _{12}+\hat{\cal B} _{13}+\hat{\cal B} _{23}\Big]
	{\cal F}^c _4({\bf r}_{12},{\bf r}_{13},{\bf r}_{23},R) = \tilde {E}^{(0)} (R)
+
	\tilde {E}^{(1)}({\bf r}_{12},{\bf r}_{13},{\bf r}_{23},R),
\label{3points}
\end{equation}
where now the function $\tilde {E}({\bf r}_{12},{\bf r}_{13},{\bf r}
_{23},R)$ is given by  equation (\ref{Eex}) but with the sum on
$\hat{\cal B}$ containing now one less operator. We seek solutions that
are symmetric with respect to permuting the pairs $12,\ 13,\
23$. The leading inhomogeneous solution can be found as before,
cf.(\ref{inhF4f}),
\begin{equation}
	{\cal F} _{4,\rm inh}^{(1)}({\bf r}_{12},{\bf r}_{13},{\bf r}_{23},R) \sim
	S_4(R){r _{12}^{\zeta_2} +r _{13}^{\zeta_2}+r _{23}^{\zeta_2}
	\over R^{\zeta_2}} \  .
\label{3inF4}
\end{equation}
The homogeneous solutions will be of the form (\ref{Fsumharm}),
but with additional summations over $l''$ and $m''$ for the
third factor $g_{l''m''}$ as a function of the third small
vector distance. This is all the information needed for the
calculations that follow.
\subsection{Calculation of K(R)}

According to our strategy we need now to evaluate the limits
shown in Eq.(\ref{limK}). We can firstly compute the derivatives
operating on the products of ${\cal F}_2$ and discover that the
limits are not singular. In fact that contributions decays
rapidly in $R$ like $R^{-2\zeta_h}$.  In computing the
derivatives of the cumulant of ${\cal F}_4$  we need to consider all
the contributions to the solution of ${\cal F} _4$  for two pairs of
coalescing points that were examined in subsection \ref{solutions},
and seek the leading one. The solution
(\ref{inhF4}) does not survive the derivative in
Eq.(\ref{limK}). The $l=l'=0$ term in (\ref{Fsumharm}) is
constant and also does not survive. The next term that survives
the derivatives is the $l=l'=2$ term, and it is less singular
than the inhomogeneous solutions (\ref{inh2F4}) since $\beta_2 >
\zeta_2$. Accordingly the leading contribution is
(\ref{inh2F4}), and upon performing the derivatives we get
\begin{equation}
	K(R) \sim {S_4(R)\over R^{2\zeta_2}}
	\lim_{r_{12},r_{34}\to 0}{\kappa^2\over
	(r_{12}r_{34})^{\zeta_h}} \ . \label{solKR}
\end{equation}
where we have used the fact that $\zeta_h = 2-\zeta_2$.  The
singular limit has to be understood in light of the full
equation for ${\cal F}^c_4$, Eq.(\ref{cumF}), in which the
$\kappa$-diffusive terms are explicit. The role of these terms
is precisely to truncate the divergence that is implied by
(\ref{solKR}). As a consequence the divergence is only
applicable in the inertial range with $r_{12},r_{34}> \eta$,
whereas in the dissipative regime the divergence disappears.
Thus for evaluating $K(R)$ via inertial range values we must
replace the  limit $r_{12},r_{34}\to 0$ by
$r_{12}=r_{34}=\eta$:
\begin{equation}
	K(R) \sim {S_4(R)\over R^{2\zeta_2}}{\kappa^2\over
	\eta^{2\zeta_h}} \ , \label{fsolKR}
\end{equation}
Comparing to Eq.(\ref{KR}) we see that we recover the result
that $\Delta=\Delta_c=\zeta_h$.

We can rewrite (\ref{fsolKR}) in a final form by introducing
$\bar\epsilon$ as
\begin{equation}
	\bar\epsilon = \kappa\left<|\nabla T|^2\right> =
	-\kappa\lim_{r_{12}\to\eta}\nabla_1
	\nabla_2{\cal F}(r_{12})\propto \kappa/\eta^{\zeta_h} \ ,
\label{eps}
\end{equation}
where we used the fact that ${\cal F}(r_{12})\sim r_{12}^{\zeta_2}$.
Using the last equation in the preceding one we find the final
result:
\begin{equation}
K(R) \simeq \bar\epsilon^2 S_4(R)/S_2(R)^2 \ . \label{finalKR}
\end{equation}
\subsection{Calculation of the correlation functions $L_{ll',m}$}

In this subsection we turn to the calculation of correlation
functions that expose the scaling properties of anomalous fields
with other irreducible  representations of the rotation group.
We can do this by introducing the following correlation
functions:
\begin{eqnarray}
	&&L_{l,l',m}(r_{12},r_{34},R) =
	\int d\cos\theta_{12} d\cos\theta_{34} d(\phi_{12}-\phi_{34})
\nonumber\\
&\times&{\cal F} _4 ({\bf r}_{12},{\bf r}_{34},R)Y_{l,m}(\theta_{12},\phi_{12})
	Y_{l',-m}(\theta_{34},\phi_{34}) \ .
\label{defL}
\end{eqnarray}
For $l=l'=0$ the leading contribution to this correlation
function arises from (\ref{inhF4f}), and
\begin{equation}
	L_{0,0,0}(r_{12},r_{34},R) \simeq
	{\cal F}^{(2)}_{4,\rm inh}\propto R^{\zeta_4-2\zeta_2} \ .
\label{L000}
\end{equation}
For $l,l' \geq 2$ the leading contribution arises from
(\ref{Fsumharm}). Using  (\ref{fllmm}) we write the final result
\begin{equation}
	L_{l,l',m}(r_{12},r_{34},R) \sim S_4(R)
	\left({r_{12}\over R}\right)^{\beta_l}
	\left({r_{34}\over R}\right)^{\beta_l'} \ . \label{Lllm}
\end{equation}
Finally, for $l=0$ and $l'\geq 2$ or vice versa the leading
contribution stems from the solution (\ref{F4(3)}). The
expression for $L_{0,l',0}(r_{12},r_{34},R)$ can be obtained
from (\ref{Lllm}) simply by replacing $\beta_0 = 0$ by $\zeta_2$.

Eq.(\ref{Lllm}) gives rise to a set of anomalous local fields.
Taking gradients $\bbox{\nabla}_{1\alpha} \bbox{\nabla}_{2\beta}$,
one produces one scalar field $(\bbox{\nabla}_1 \cdot
\bbox{\nabla}_2)$, which corresponds to $l = 0$ and has anomalous
scaling $\zeta_h = 2 - \zeta_2$, and the traceless tensor
$(\bbox{\nabla}_{1\alpha} \bbox{\nabla}_{2\beta} - \frac{1}{3}
\bbox{\nabla}_1 \bbox{\nabla}_2)$ corresponding to $l = 2$.
Taking four gradients one can produce anomalous fields originating
from $l = 4$, and so on.

\subsection{Calculation of the cumulant $J_4^c(R)$}

The quantity $J_4(R)$ has reducible contributions which were
computed above, and a cumulant part which is obtained from ${\cal F}
^c_4$.  The calculation of  $J_4^c(R)$ follows very much the
lines of the calculation of $K(R)$, except that one needs to
find again which of the solutions found in sec\ref{solutions}
contributes mostly to equations (\ref{limJ1}) and (\ref{limJ2}).
It turns out that the leading contribution to  (\ref{limJ1})
stems from the solution (\ref{inhF4}), whereas the leading
contribution to (\ref{limJ2}) arises from the solution
(\ref{3inF4}). As before the derivatives with respect to $r_1$
and $r_2$ result in a singular limit  when $r_{12}\to 0$, and we
have to take $r_{12}=\eta$ after computing the derivatives. On
the other hand the other limits are regular and they do not
require special care. The evaluation of $J_{4,1}(R)$ and
$J_{4,2}(R)$ turn out to be the same up to constants. The result
is
\begin{equation}
J_4^c(R) \sim \kappa S_4(R)/R^{\zeta_2}\eta^{\zeta_h} \ . \label{J4nf}
\end{equation}
Using Eq.(\ref{eps}) this can be written finally as
\begin{equation}
J_4^c(R) = C_4 \bar\epsilon S_4(R)/S_2(R) \ , \label{J4f}
\end{equation}
where $C_4$ is a dimensionless constant that will be determined
below. Together with the reducible contributions that were
computed above we can write
\begin{equation}
	J_4(R) = \bar\epsilon S_2(R) + {\kappa\over 2}
	[\nabla S_2(R)]^2 + C_4 \bar\epsilon S_4(R)/S_2(R) \ .
\label{finalJ4}
\end{equation}
It is obvious that the second term is small compared with the
first and it can be neglected.

\section{The calculation of $J_{2n}(R)$}.

The correlations (\ref{J}) which make $J_{2n}(R)$ have Gaussian
decompositions and a cumulant part. The leading contribution to
the Gaussian decomposition is
\begin{equation}
	J_{2n}^G(R) = \bar\epsilon S_{2n-2}(R)
\end{equation}
which corresponds to the first term in (\ref{finalJ4}). To
evaluate the cumulant part one needs to compute correlation
functions of the type
\begin{equation}
	J_{p,q}(R) = \left<\left<|\nabla_1\Theta({\bf r}_1)|^2
	\Theta^{p-2} ({\bf r}_1)\Theta^q ({\bf r}_1)\right>\right> \ .
\label{Jpq}
\end{equation}
with ${\bf r}_1=-{\bf R} /2$ and ${\bf r}_2 = {\bf R} /2$.
The calculation of $J_{p,q}(R)$ follows from the equation for
${\cal F}^c_{2n}$ upon coalescing a group of $p$ points (denoted
below as the $\alpha$ group) into the position $-{\bf R} /2$ and a
group of $q$ points (denoted as the $\beta$ group) into the
position ${\bf R}/2$. We start with Eq.(\ref{3.28}) which in the
inertial interval takes on the form
\begin{equation}
	\Big[\sum_{\alpha>\alpha'=1}^p \hat{\cal B}_{\alpha\alpha'} +
	\sum_{\beta>\beta'=p+1}^{2n} \hat{\cal B}_{\beta\beta'}
	+\sum_{\alpha = 1}^p \sum_{\beta=p+1}^{2n}
	\hat{\cal B}_{\alpha\beta}\Big]{\cal F}^c_{2n}({\bf r}_1,\dots,{\bf r}_{2n})=
RHS \ .
\label{EqF2n}
\end{equation}
As before the effective equation is obtained by grouping
together a quantity $E(\{r_{\alpha\alpha'}\},
\{r_{\beta\beta'}\},R) = RHS - \sum\sum
\hat{\cal B}_{\alpha\beta}{\cal F}^c_{2n}({\bf r}_1,\dots,{\bf r}_{2n})$. To
find
the form of the solution we note that when we had one pair of
coalescing points this led to the solution (\ref{Fin}). Two
pairs of coalescing points led to (\ref{inhF4f}), whereas three
pairs of coalescing points resulted in (\ref{3inF4}). In the
present case we have $[p(p-1)+q(q-1)]/2$  coalescing pairs and
the solution which belongs to the same family is written as
\begin{equation}
	{\cal F}_{2n,\rm inh} \sim S_{2n}(R)
	\Big[\sum_{\alpha>\alpha'=1}^p r_{\alpha\alpha'}^{\zeta_2}
 	+ \sum_{\beta>\beta'=p+1}^{2n} r_{\beta\beta'}^{\zeta_2}\Big]
 	/R^{\zeta_2} \ . \label{F2nin}
\end{equation}
Next we need to compute the derivative with respect to ${\bf r}_1$
and ${\bf r}_2$ and take the limit of all pairs of coalescing
distances going to zero. The only divergence will be associated
with $r_{12}^{-\zeta_h}$ whereas all the other limits are
trivial. According to our strategy we have to  cut the
divergence at $r_{12}=\eta$ and thus we calculate ${\cal F}_{2n,\rm
inh}\sim S_{2n}(R)/\eta^{\zeta_h}$. Using again Eq.(\ref{aa}) we
have
\begin{equation}
	J_{2n}^c(R) = C_{2n}\bar{\epsilon} S_{2n}(R)/S_2(R) \ ,
\label{J2nf}
\end{equation}
where $C_{2n}$ is an unknown dimensionless coefficient that will
be determined soon. This equation is the generalization of
(\ref{J4f}) for any $n$. Combining now the cumulant with the
leading reducible contribution we get the final result
\begin{equation}
	J_{2n}(R) = \bar\epsilon \Big[S_{2n-2}(R)+C_{2n}
	S_{2n}(R)/S_2(R)\Big] \ . \label{J2nfinal}
\end{equation}

\section{Scaling exponents of the structure functions}

In this section we collect all the results obtained above with
the aim of reaching  conclusions about the scaling exponents of
the structure functions. The first result that we need to pay
attention to is (\ref{fsolKR}) or (\ref{finalKR}) for $K(R)$.
This result  shows that $K(R)\sim R^{\zeta_4-2\zeta_2}$. Since
$K(R)$ cannot be an increasing function of $R$ we conclude
immediately that
\begin{equation}
\zeta_4-2 \leq 2\zeta_2 \ . \label{expineq}
\end{equation}
Next we need to consider (\ref{finalJ4}) for $J_4(R)$, rewriting
it after neglecting the second term as
\begin{equation}
	J_4(R) \simeq \bar\epsilon S_2(R)\Big[1+C_4
	{S_4(R)\over S_2^2(R)}\Big] . \label{J4RR}
\end{equation}
The second term in the parenthesis is dimensionless, and may be
written as $(\ell/R)^{2\zeta_2-\zeta_4}$.

Now there are two possibilities:

(i) $\zeta_4=2\zeta_2$ and the scaling is normal. The first and
third terms have the same scaling and they are of the same
order.

(ii) ${\zeta_4 < 2\zeta_2}$, and the scaling is anomalous. If
so, the second term in (\ref{J4RR}) must be larger than the
first. For that to happen the renormalization scale {\it must}
be the outer scale $L$.

The implications of the first possibility are somewhat strange.
For example if there is normal scaling the correlation function
$K(R)$ does not decay in $R$. This means that the dissipation
field is not mixing. On the contrary, if there is anomalous
scaling, then the correlation $K(R)$ decays as is expected from
a random field. We will explore now the second possibility and
show that if there is anomalous scaling then the scaling
exponents can be computed in agreement with Kraichnan's
arguments.

If we accept anomalous scaling, then the leading term in
(\ref{J2nfinal}) is the second one. Generally speaking
$J_{2n}(R)$ may be written (following Kraichnan et al.
\cite{95KYC}) as
\begin{equation}
	J_{2n}(R) = 4n\kappa\int d\Delta\Theta P(\Delta\Theta)
	[\Delta\Theta]^{2n-2}
	\left<|\nabla\Theta|^2 |\Delta\Theta\right> \ , \label{cond}
\end{equation}
where $\left<|\nabla\Theta|^2 |\Delta\Theta\right>$ is the
average of $|\nabla\Theta|^2$ {\it  conditional} on a given
value of $\Delta\Theta$. $P(\Delta\Theta)$ is the probability to
observe     a given value of $\Delta\Theta$. Eq.(\ref{cond}) is
exact. The question now is what is the dependence of
$\left<|\nabla\Theta|^2 |\Delta\Theta\right>$ on $\Delta\Theta$.
In order to recover our result (\ref{J2nf}) this conditional
average {\it must} satisfy
\begin{equation}
	\left<|\nabla\Theta|^2 |\Delta\Theta\right> =
	C (\Delta\Theta)^2/S_2(R). \label{amaz}
\end{equation}
This means that the coefficient $C_{2n}$ in (\ref{J2nf}) is
$n$-independent. We can determine this coefficient from the
particular case $n=1$. Since $J_2 = \bar{\epsilon}$ the
conclusion is that $C_n=1$. Equipped with this we consider
again the balance equation (\ref{balance})
\begin{equation}
 R^{1-d}{\partial \over \partial R} R^{d-1}h(R)
{\partial \over \partial R }S_{2n}(R) = J_{2n}(R) \ . \label{bal}
\end{equation}
Using now the definition of the scaling exponents $S_{2n}(R)\sim
R^{\zeta_{2n}}$ we retrieve from (\ref{bal}) the result
(\ref{kr_screl}) which can be also written as
\begin{equation}
	\zeta_{2n} = {1\over 2}\Bigg[\zeta_2 -d  +
	\sqrt{(\zeta_2+d)^2+4\zeta_2(n-1)} \ \Bigg] \ . \label{kra}
\end{equation}
This is Kraichnan's anomalous scaling.
\section{summary and conclusions}

The first conclusion of this paper is that renormalized
perturbation theory for hydrodynamic fields has the potential to
describe nonperturbative effects. Exact resummations of the
diagrammatic series result in exact equations for the
statistical quantities that contain not only perturbative but
also nonperturbative effects.

Secondly, the passive scalar problem with rapidly decorrelating
velocity field displays a particularly simple analytic structure
in which all the scaling properties emanate from one
differential operator $\hat{\cal B}(R)$. As a result this theory becomes
``critical" in the sense that the anomalous exponent $\Delta$
that was introduced in paper I is exactly critical. The reason
for this is that both $\Delta$ and $\Delta_c$ come from the same
operator, and thus they must be the same. In such a situation
the subcritical scenario that was suggested in \cite{95LP} is
untenable in this case.

Thirdly, the simplicity of the theory allowed us to compute the
whole spectrum of anomalous exponents that are associated with
the ultraviolet divergence. It was shown that these exponents
are related to the spectrum $\beta_l$ (\ref{betal2}) which
appears in the law of isotropization of the 2-point
correlation function. The same exponents determine the $R$
dependence of the correlation functions $L_{ll',m}$ that were
introduced in (\ref{defL}).

Finally we showed that as far as the structure functions
$S_n(R)$ are concerned, there are only two possibilities. Either
the exponents satisfy $\zeta_{2n} =n\zeta_2$ and the scaling is
normal or  the scaling is anomalous with the outer
renormalization scale and with the law (\ref{kra}). We noted
that normal scaling also implies that the dissipation field is
not mixing.

Whether or not the ``critical" situation of this model with a
marginal scaling exponent $\Delta= \Delta_c$  is structurally
stable is an extremely important question that must await
further research.

\acknowledgments
Numerous illuminating discussions with Bob Kraichnan were
influential on our thinking and are  gratefully acknowledged.
This work was supported in part by the German-Israeli
Foundation, the Basic Research Fund of the Israel Academy of
Sciences and the N. Bronicki fund.

\appendix

\section{Derivation of the equation for the simultaneous two-point
correlator}

Here we obtain the equation of motion for the simultaneous two-point
correlator,
${\cal F}({\bf r}_1,{\bf r}_2,t=0)$.
We make use of Eq.
(\ref{g_eq}) to verify that the operator

\begin{equation}
   	{\cal G}_{12}^{-1}({\bf r}_0|{\bf r}_1,{\bf r}_2,t) \equiv \partial_t +
      	\hat{\cal D}_1({\bf r}_1 - {\bf r}_0) + \hat{\cal D}_1({\bf r}_2 - {\bf
r}_0)
\end{equation}
is the inverse of the product of Green's functions,
that is,
\begin{eqnarray}
 \label{g12_inv}
& & 	{\cal G}_{12}^{-1}({\bf r}_0|{\bf r}_1,{\bf r}_2,t)  {\cal G}({\bf
r}_0|{\bf r}_1,{\bf r}_1',t)
	{\cal G}({\bf r}_0|{\bf r}_2,{\bf r}_2',t) \nonumber \\
& & =
	\delta({\bf r}_1-{\bf r}_1')\delta({\bf r}_2-{\bf r}_2') \delta(t).
\end{eqnarray}
We have used here the fact that the Green's function when
multiplied by $\delta(t)$ may be evaluated at $t=0$. Applying
this operator and performing the spatial integrations, one
obtains directly
\begin{equation}
\label{f_zerot}
     	\left[\hat{\cal D}_1({\bf r}_1-{\bf r}_0) +
     		\hat{\cal D}({\bf r}_2-{\bf r}_0) \right]
	{\cal F}({\bf r}_1,{\bf r}_2) = \Phi({\bf r}_0|{\bf r}_1,{\bf r}_2) +
\Phi_0({\bf r}_1-{\bf r}_2).
\end{equation}
Now using the definition of $\Phi({\bf r}_0|{\bf r}_1,{\bf r}_2)$ from
Eq.(\ref{phi}) and dropping the ${\bf r}_0$ dependence, one may rewrite this
as
\begin{equation}
	-\left[ \kappa(\nabla^2_1 + \nabla^2_2) +
	h_{ij}({\bf r}_1-{\bf r}_2) \frac{\partial}{\partial r_{1i}}
	 \frac{\partial}{\partial r_{2j}} + \hat{{\cal H}}({\bf r}_1,{\bf r}_2)
	 \right] {\cal F}({\bf r}_1,{\bf r}_2) = \Phi_0({\bf r}_1,{\bf r}_2),
\end{equation}
where the operator $\hat{{\cal H}}$ is given by
\begin{equation}
	\hat{{\cal H}}({\bf r}_1,{\bf r}_2) \equiv
	\left( h_{ij}({\bf r}_1) \frac{\partial}{\partial r_{1i}}
	 + h_{ij}({\bf r}_2) \frac{\partial}{\partial r_{2i}} \right)
	 \left(\frac{\partial}{\partial r_{1j}} +
	 \frac{\partial}{\partial r_{2j}} \right).
\end{equation}
In the case that all quantities are only functions of ${\bf R} =
{\bf r}_1 - {\bf r}_2$, $\hat{{\cal H}}$ vanishes. The remaining terms
are equivalent to the definition of the operator
$\hat{\cal D}_2({\bf R})$ in Eq.(\ref{def-D_2}) and one recovers
Eq.(\ref{sim-F}).

\section{Equation for the cumulant of the $2n$-point correlator}

In this appendix we derive equations for the many-time
correlators using the diagrammatic approach.  The
infinite diagrammatic series for the 4-point
correlator was presented in \cite{94LPF}. Let us
consider  the series for the correlator.  A typical
diagram for a $2n$-point correlator consists of $n$
'tramways': strings of Green's functions, connected in
pairs at one end by two-point correlators. These
tramways may be interconnected by velocity
correlators. Therefore to build up an arbitrarily
complex diagram, one adds successively connections
between chosen pairs of tramways. Let us consider a
given pair of end-points, $x_\alpha$ and $x_\beta$.
Group all diagrams together in which the Green's
function beginning at $x_\alpha$ and that beginning at
$x_\beta$ are linked by a velocity correlator. The
series of diagrams to the right of this first velocity
correlator is again the series of
diagrams for the full correlator.

The difficulty with writing a resummed equation for the
full correlator is that the lowest order terms are
Gaussian, and disconnected. In building upon the
Gaussian terms by the addition of `rungs' as just
described, one will generate, amongst other terms,
a series in which the disconnected parts remain
disconnected, and may be resummed again into the Gaussian
decomposition. This means that these terms actually
appear twice. In order to avoid this one may write an
an equation for the {\em cumulants}. For the cumulant
the lowest order terms are those in which all tramways
have one velocity correlator connection to another.

To simplify  the appearance of this equation we introduce
the operator  $\widehat {C}_{\alpha \beta}$ which
represents the addition of a rung, and here operates on
${\cal F}_{2n} (0|x_1,...,x_\alpha,x_\beta,...,x_{2n})$.
The definition is
\begin{eqnarray}
	\widehat {C}_{\alpha \beta} &*&
	{\cal F}_{2n}(0|x_1,...,x_\alpha,x_\beta,...,x_{2n})
	\nonumber   \\
& \equiv  &
  	\int d{\bf r}_1' d{\bf r}_2'\int_{t_m}^\infty  dt
	{\cal G}^0_2
	(0|x_\alpha,x_\beta,{\bf r}_1',t,{\bf r}_2',t)
\label{3.8}\\
&\times&
	H_{ij}({\bf r}_1',{\bf r}_2')
  	\frac{\partial}{\partial r'_1}
  	\frac{\partial}{\partial r'_2}
  	{\cal F}_{2n}
  	  	(0|x_1,...,{\bf r}_1',t,{\bf r}_2',t,...,x_{2n})  .
\nonumber
\end{eqnarray}
Since ${\cal F}_{2n}(0| x_1,x_2,..,x_{2n})$ is symmetric
with respect to all  exchanges of coordinates this
definition is sufficient for any pair of indices $1 \leq
\alpha, \beta  \leq 2n$.

The lowest order terms may be expressed in terms
of the operator $\hat{C}$ and two-point correlators as
\begin{equation}
	{\cal F}_{2n,0}^c(0|x_1,x_2,..,x_{2n}) =
	\prod_{j=2,4}^{2n} \hat{C}_{j,j+1} * \prod_{k=1,3..}^{2n-1}
	{\cal F}(x_k,x_{k+1})
\end{equation}
Then the resummed equation has
the form
\begin{eqnarray}
	{\cal F}_{2n}^c(0|x_1,x_2,..,x_{2n}) & = &
	\sum_{\rm perm} {\cal F}_{2n,0}^c(0|x_1,x_2,..,x_{2n})
\nonumber \\
	& + & \sum_{\alpha>\beta} \hat{C}_{\alpha \beta} *
	{\cal F}_{2n}^c(0|x_1,x_2,..,x_{2n}).
\end{eqnarray}
Now operate on both sides of the equation with the product
of the inverse Green's functions (\ref{g_inv}).
Using the fact that
\begin{equation}
	\left( \frac{\partial}{\partial t_\alpha} +
	\hat{\cal D}_1({\bf r}_\alpha) \right)
	\left( \frac{\partial}{\partial t_\beta}
		+ \hat{\cal D}_1({\bf r}_\beta) \right)
	C_{\alpha \beta} = \delta(t_\alpha - t_\beta)
	{\cal B}_{\alpha \beta},
\end{equation}
one finds
\begin{eqnarray}
& & \prod_{k = 1}^{2n}
	\left( \frac{\partial}{\partial t_k} +
	\hat{\cal D}_1({\bf r}_k) \right)
	{\cal F}_{2n}^c(0|x_1,x_2,..,x_{2n})
=  \nonumber \\
& &	\sum_{\left<\alpha, \beta \right>}
	\sum_{\scriptstyle(\gamma_i)
	\atop{\rm perm}(\gamma_1 \neq \alpha, \beta)}
	\left[\prod_{i=1}^{2n-2}
	\delta(t_{\gamma_i} - t_{\gamma_{i+1}}) \right]
	\left( \frac{\partial}{\partial t_\alpha} +
	\hat{\cal D}_1({\bf r}_\alpha) \right)
	\left( \frac{\partial}{\partial t_\beta} +
		\hat{\cal D}_1({\bf r}_\beta) \right)
	{\cal B}_{\gamma_1,\gamma_2}
	{\cal F}(x_\alpha,x_{\gamma_1})
\nonumber \\
& &	\left\{ \prod^{2n-4}_{j=2,4..}
	{\cal B}_{\gamma_{j+1},\gamma_{j+2}}
	{\cal F}({\bf r}_{\gamma_{j}},{\bf r}_{\gamma_{j+1}})
	\right\}
	{\cal B}_{\gamma_{j+1},\gamma_{j+2}}
	{\cal F}(x_{\gamma_{2n-2}},x_\beta)
\nonumber \\
& &	+  \sum_{\alpha > \beta} 	\delta(t_\alpha-t_\beta)
	\prod_{\gamma \neq \alpha,\beta}
	\left( \partial_t + \hat{\cal D}_1({\bf r}_\gamma) \right)
	{\cal B}_{\alpha \beta}
	{\cal F}_{2n}^c(0|x_1,x_2,..,x_{2n}).
\end{eqnarray}

Therefore we have derived a closed equation for the time development
of the many-time cumulants of the $2n$-point correlator in
terms of the operator ${\cal B}$ and the two-point moments only.
We will not solve this equation in this paper.




\end{document}